\documentclass{aa}  

\usepackage{graphicx}
\usepackage{color}
\usepackage{txfonts}
\begin{document}

   \title{Signatures of the non-Maxwellian $\kappa$-distributions \\in optically thin line spectra}
   \subtitle{II. Synthetic Fe XVII--XVIII X-ray coronal spectra and predictions \\for the \textit{Marshall Grazing-Incidence X-ray Spectrometer (MaGIXS)}}

   \author{Jaroslav Dud\'ik\inst{1}\fnmsep\thanks{RS Newton Alumnus}
 	  \and Elena Dzif\v{c}\'akov\'a\inst{1}
         \and Giulio Del Zanna\inst{2}
	  \and Helen E. Mason\inst{2}
	  \and \\ Leon L. Golub\inst{3}
	  \and Amy R. Winebarger\inst{4}
	  \and Sabrina L. Savage\inst{4} 
          }

   \institute{Astronomical Institute, Academy of Sciences of the Czech Republic, 25165 Ond\v{r}ejov, Czech Republic,\\
               \email{jaroslav.dudik@asu.cas.cz}
         \and Department of Applied Mathematics and Theoretical Physics, CMS, University of Cambridge, Wilberforce Road,
	      Cambridge CB3 0WA, United Kingdom
	 \and Harvard-Smithsonian Center for Astrophysics, 60 Garden Street, Cambridge MA 01238, USA
         \and NASA Marshall Space Flight Center, Mail Code ST 13, Huntsville, AL 35812, USA\\
             }

   \date{Received ; accepted }

  \abstract
   {}
   {We investigated the possibility of diagnosing the degree of departure from the Maxwellian distribution using the \ion{Fe}{XVII}--\ion{Fe}{XVIII} spectra originating in plasmas in collisional ionization equilibrium, such as in the cores of solar active regions or microflares.}
   {The original collision strengths for excitation are integrated over the non-Maxwellian electron $\kappa$-distributions characterized by a high-energy tail. Synthetic X-ray emission line spectra were calculated for a range of temperatures and $\kappa$. We focus on the 6--24\,\AA~spectral range to be observed by the upcoming Marshall Grazing-Incidence X-ray Spectrometer \textit{MaGIXS}.}
   {We find that many line intensity ratios are sensitive to both $T$ and $\kappa$. Best diagnostic options are provided if a ratio involving both \ion{Fe}{XVII} and \ion{Fe}{XVIII} is combined with another ratio involving lines formed within a single ion. The sensitivity of such diagnostics to $\kappa$ is typically a few tens of per cent. Much larger sensitivity, of about a factor of two to three, can be obtained if the \ion{Fe}{XVIII} 93.93\,\AA~line observed by \textit{SDO}/AIA is used in conjuction with the X-ray lines.}
   {We conclude that the \textit{MaGIXS} instrument is well-suited for detection of departures from the Maxwellian distribution, especially in active region cores.}

   \keywords{Sun: UV radiation -- Sun: X-rays, gamma rays -- Sun: corona -- Radiation mechanisms: non-thermal}
   \maketitle
%
\section{Introduction}
\label{Sect:1}

The coronal heating problem \citep[e.g.,][]{Klimchuk06,Klimchuk15,Schmelz15a} has been with us for many decades. Highest plasma temperatures (outside solar flares) are observed within the cores of active regions, where the temperatures reach 3\,MK or more. These locations emit prominently in soft X-ray and extreme-ultraviolet (EUV) wavelengths \citep[e.g.,][]{Parkinson73,Parkinson75,Walker74,Hutcheon76,Bromage77,Saba91,Brosius96,Nagata03,Reale10,Winebarger11,Warren11,Warren12,Teriaca12a,Testa12,DelZanna13b,DelZanna14a,DelZanna15b,Nuevo15,Schmelz15b,Li15,Brooks16,Ugarte-Urra17,Parenti17}. Emission at such temperatures are also observed from stars with coronae, such as Procyon \citep[e.g.,][]{Testa12}, including active stars, such as Capella \citep[e.g.,][]{Phillips01,Desai05,Clementson13,Beiersdorfer11,Beiersdorfer14b}, $\alpha$\,Cen\,B, AB\,Dor, and others \citep[e.g.,][]{Wood18}.

The emission of the solar active region cores is typically observed in spectral lines such as \ion{Fe}{XVI}--\ion{Fe}{XVIII} \citep[see][for a recent review]{DelZanna18}, or within X-ray or EUV filters such as those onboard \textit{Hinode}/XRT \citep{Golub07} or the 94\,\AA~channel of the Atmospheric Imaging Assembly \citep[AIA][]{Lemen12,Boerner12} onboard \textit{Solar Dynamics Observatory}. Emission originating at temperatures even higher than 3\,MK can also be present, as a consequence of impulsive (nanoflare) energy release \citep{Reale09a,Reale09b,Schmelz09a,Schmelz09b,Patsourakos09}, that likely recurs at timescales comparable to cooling timescales of individual coronal loops \citep[e.g.,][]{Cargill14,Bradshaw16}. The emission measure of plasma at such temperatures is however likely low \citep[see, e.g.,][]{Warren12}, and in addition, the current instrumentation has a ``blind spot'' at higher temperatures \citep{Winebarger12b}. Upper limits to the emission at higher temperatures are provided by \citet{DelZanna14a} using earlier X-ray spectra from \textit{SMM}/FCS and by \citet{Parenti17} using ultraviolet \textit{SoHO}/SUMER spectra.

To overcome this difficulty and to observe plasma that is typically formed in the active region cores, as well as to elucidate the properties of its heating, observations in the X-ray spectral range are needed. Such observations should be provided by the upcoming \textit{Marshall Grazing-Incidence X-ray Spectrometer} \citep[\textit{MaGIXS}][]{Kobayashi10,Kobayashi17,Kobayashi18}. \textit{MaGIXS} is an X-ray spectrometer designed to observe the solar spectrum at 6--24\,\AA~(0.5--2.0\,keV) with a spectral resolution better than 50 m\AA~near the center of a slit of $\pm$4$\arcmin$ length. Its spatial resolution is about 5$\arcsec$. The launch of \textit{MaGIXS} on a NASA sounding rocket is currently planned for spring 2020. On the basis of previous X-ray observations \citep[cf.,][]{Parkinson75,DelZanna14a}, we expect \ion{Fe}{XVII} lines to be prominent in \textit{MaGIXS} spectra. For active conditions, \ion{Fe}{XVIII} and perhaps higher ionization stages could be present. We therefore have considered here \ion{Fe}{XVII} and \ion{Fe}{XVIII} our primary diagnostic ions. The recent atomic data and line identifications for these ions have been reviewed by \citet{DelZanna06,DelZanna11b}. 

An additional science objective of \textit{MaGIXS} is to elucidate the possible presence of departures from the equilibrium Maxwellian distribution of electrons in the plasma of active regions. The non-Maxwellians, especially in the form of high-energy tails \citep[see review of][]{Dudik17b}, could arise as a consequence of energy release (including impulsive heating), whether by reconnection \citep[e.g.,][]{Cargill12,Gordovskyy14} including merging of magnetic islands \citep[][with the power-law tail depending on reconnection parameters]{Drake13,Montag17}; wave-particle interaction \citep[e.g.,][]{Petrosian04,Laming07,Vocks08,Vocks16}, turbulence \citep[e.g.,][]{Hasegawa85,Che14,Bian14}, or combination of these processes, such as excitation of waves by impulsive energy release, and subsequent non-local wave dissipation \citep[see, e.g., Sect. 4 of][for an estimate of wave dissipation timescales]{Arregui15}. Alternatively, high-energy tails arise wherever the ratio of the electron mean-free path to the pressure scale-length (the Knudsen number) is less than about 0.01 \citep{Scudder13}, a condition that is expected to commonly occur in solar and stellar coronae. The fundamental reason for the presence of high-energy tails is the behavior of electron cross-section for Coulomb collisions, which behaves as $E^{-2}$, leading to the collisonal frequency $\tau_\mathrm{coll}$ scaling as $E^{-3/2}$, where $E$ is the electron kinetic energy. This means that the progressively higher-$E$ electrons are more difficult to equilibrate. In fact, the high-energy tail can take hundreds or thousands times longer to equilibrate than the thermal core of the distribution \citep{Galloway10}.

While non-Maxwellian distributions containing high-energy, power-law tails are routinely detected in solar wind \citep[e.g.,][]{Marsch82a,Marsch82b,Maksimovic97a,Maksimovic97b,LeChat10,LeChat11}, in solar flares \citep[e.g.,][]{Seely87,Kasparova09,Veronig10,Fletcher11,Oka13,Oka15,Battaglia13,Battaglia15,Simoes15,Kuhar16}, and can even be present in microflares \citep[][]{Hannah10} as a high-energy excess at energies of $\gtrsim$\,5\,keV \citep{Glesener17,Wright17}, their detection elsewhere remains difficult. For example, only upper limits can be obtained in some microflare events in both active regions and quiet Sun \citep{Marsh17,Kuhar18} even with current X-ray instrumentation such as the \textit{Nuclear Spectroscopic Telescope Array} \citep[\textit{NuSTAR},][]{Harrison13} and \textit{Focusing Optics X-ray Solar Imager} \citep[\textit{FOXSI},][]{Glesener16}. 

Detection of non-Maxwellians from line spectra has also been investigated, however it is non-trivial as well. To detect departures from a Maxwellian, ratios of intensities of two lines of the same ion must either have sufficiently different excitation energy tresholds or different behavior of the excitation cross-section with energy. \citet{Dudik14b} investigated the sensitivity of the EUV lines of \ion{Fe}{IX}--\ion{Fe}{XIII} observed by the Extreme-Ultraviolet Imaging Spectrometer \citep[EIS,][]{Culhane07} onboard the \textit{Hinode} satellite \citep{Kosugi07}. It was found that a vast majority of single-ion line ratios show almost no sensitivity to the non-Maxwellian $\kappa$-distributions \citep[e.g.,][]{Owocki83,Livadiotis09,Livadiotis15,Livadiotis17}, even for strong non-thermal tails (i.e., low $\kappa$). This comes from the fact that most EIS lines of \ion{Fe}{IX}--\ion{Fe}{XIII} have similar wavelengths and are formed from similar energy levels. The temperature-sensitive ratios were an exception, showing a sensitivity of the order of several tens of per cent \citep{Dudik14b}. Typically, such ratios involved one line from the short-wavelength channel, while the other was from the long-wavelength channel of EIS. Such ratios, in combination with ratios involving lines from neighboring ionization stages, were subsequently used by \citet{Dudik15} to detect indications of strong departures from a Maxwellian ($\kappa$\,$\lesssim$\,2) in a transient coronal loop. However, such analysis is complicated by the different in-flight degradation of the two EIS wavelength channels \citep{DelZanna13a}. \citet{Dzifcakova18} used the line ratios of \ion{Fe}{XVIII}--\ion{Fe}{XIX} and \ion{Fe}{XXI}--\ion{Fe}{XXII} observed by the Extreme-Ultraviolet Experiment \citep[EVE,][]{Woods12} onboard the \textit{SDO} during an X-class flare. Lines formed at widely different wavelengths were available for diagnostics, and the results showed strong departures ($\kappa$\,$\lesssim$\,2) from the Maxwellian in the early and impulsive phase of the flare, with subsequent Maxwellianization toward the peak and gradual phases of the flare. Finally, indications of strongly non-Maxwellian distributions were found from the profiles as well as intensities of the transition region lines \citep{Dudik17a} observed by the \textit{Interface Region Imaging Spectrograph} \citep[\textit{IRIS},][]{DePontieu14}. Non-Maxwellian line profiles were also detected from EIS observations of coronal holes \citep{Jeffrey18} and flares \citep{Polito18}.

The 6--24\,\AA~passband of \textit{MaGIXS} contains numerous \ion{Fe}{XVII} and \ion{Fe}{XVIII} lines, which are the focus of this study. These lines originate in both active region cores and microflares, while the passband of \textit{MaGIXS} could contain unexplored possibilities for the detection of non-Maxwellian distributions. In addition, the \ion{Fe}{XVIII} 93.93\,\AA~line observed by AIA, widely separated in wavelength from the \textit{MaGIXS} passband, can be utilized to increase the sensitivity to both temperature and the non-Maxwellians. This work is organized as follows. The non-Maxwellian $\kappa$-distributions are described in Sect. \ref{Sect:2}. The method for calculations of synthetic spectra are detailed in Sect. \ref{Sect:3}, while the results are given in Sect. \ref{Sect:4}. A summary is given in Sect. \ref{Sect:5}. Finally, the Appendix \ref{Appendix:Transitions} provides details on the transitions (spectral lines) investigated, together with their blends and self-blends.

\begin{figure*}
   \centering
   \includegraphics[width=8.8cm]{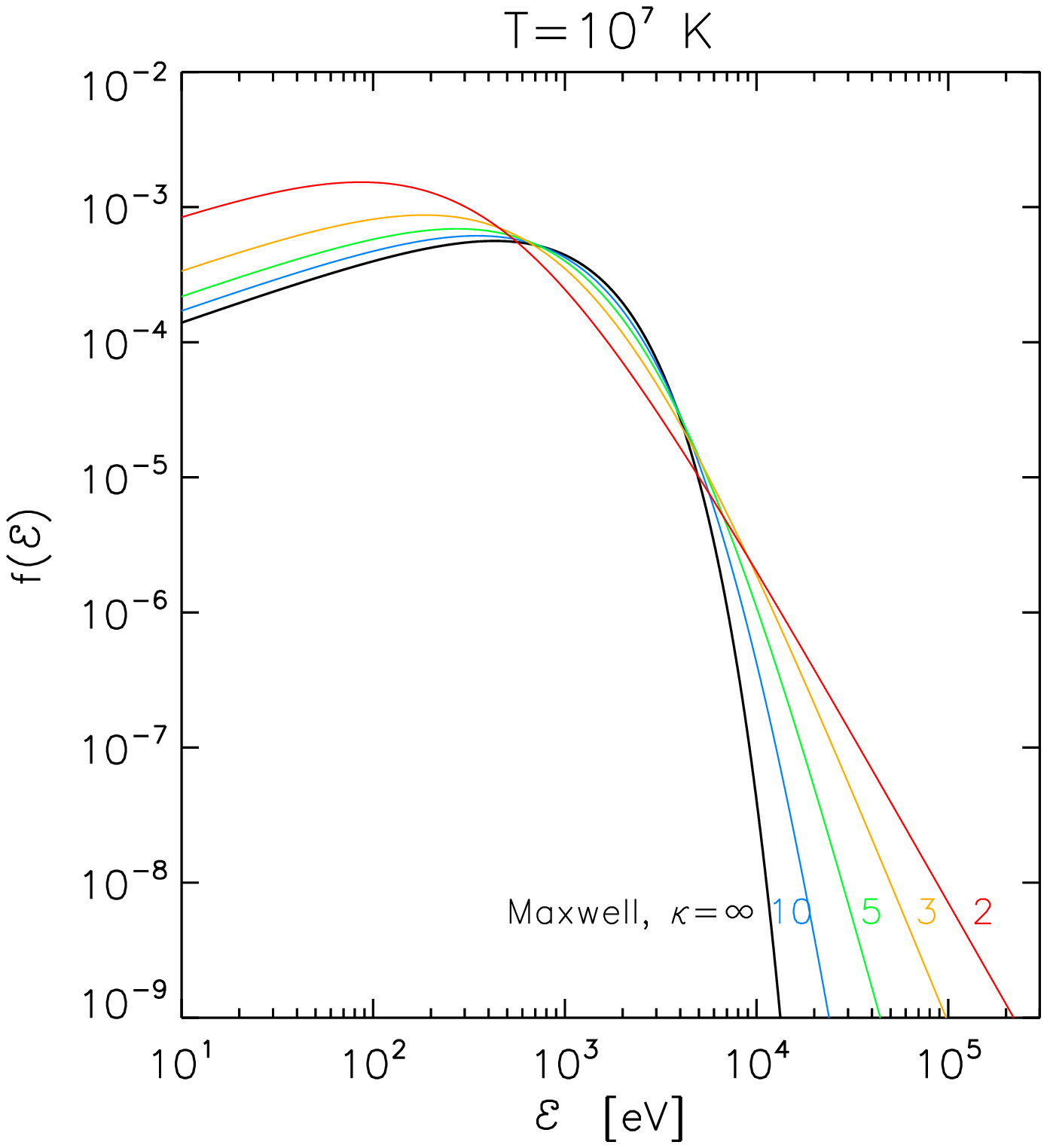}
   \includegraphics[width=8.8cm]{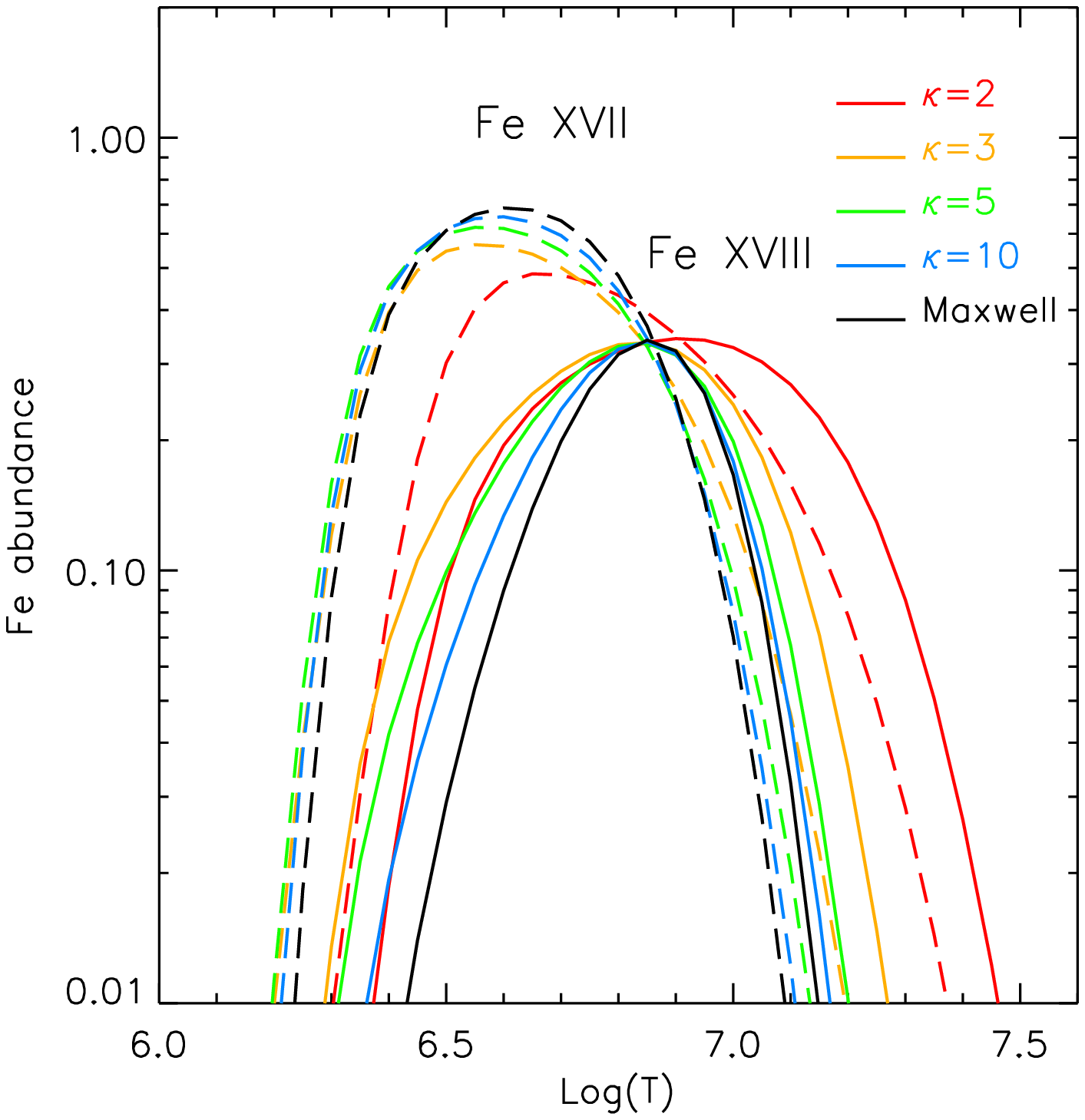}
   \caption{\textit{Left}: The electron $\kappa$-distribution as a function of electron kinetic energy $E$. \textit{Right}: Relative abundances of the \ion{Fe}{XVII} (\textit{dashed lines}) and \ion{Fe}{XVII} (\textit{full lines}). Different colors correspond to different distributions.}
   \label{Fig:ioneq}
\end{figure*}
%
%
%
\begin{figure*}
   \centering
   \includegraphics[width=8.8cm]{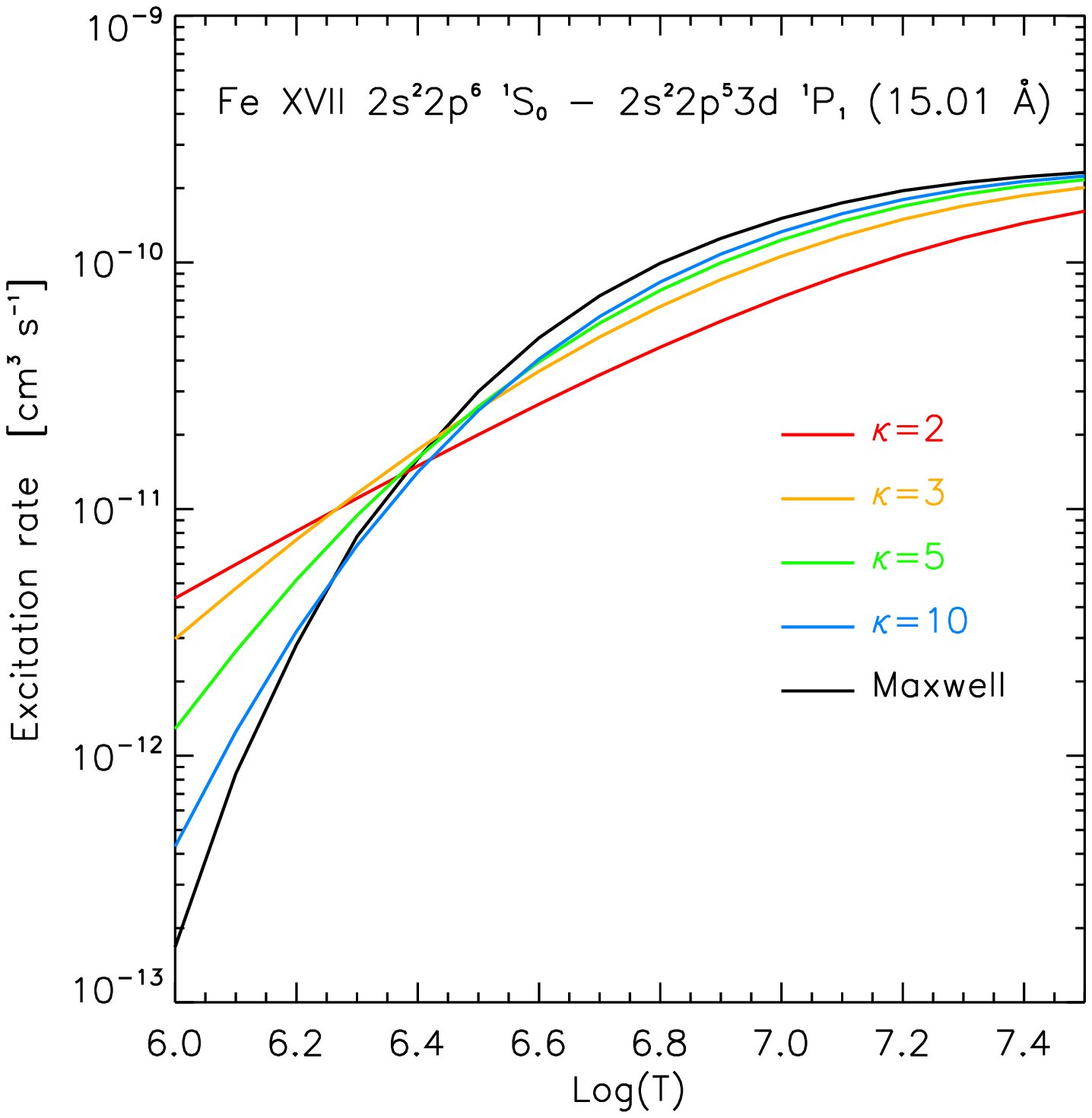}
   \includegraphics[width=8.6cm]{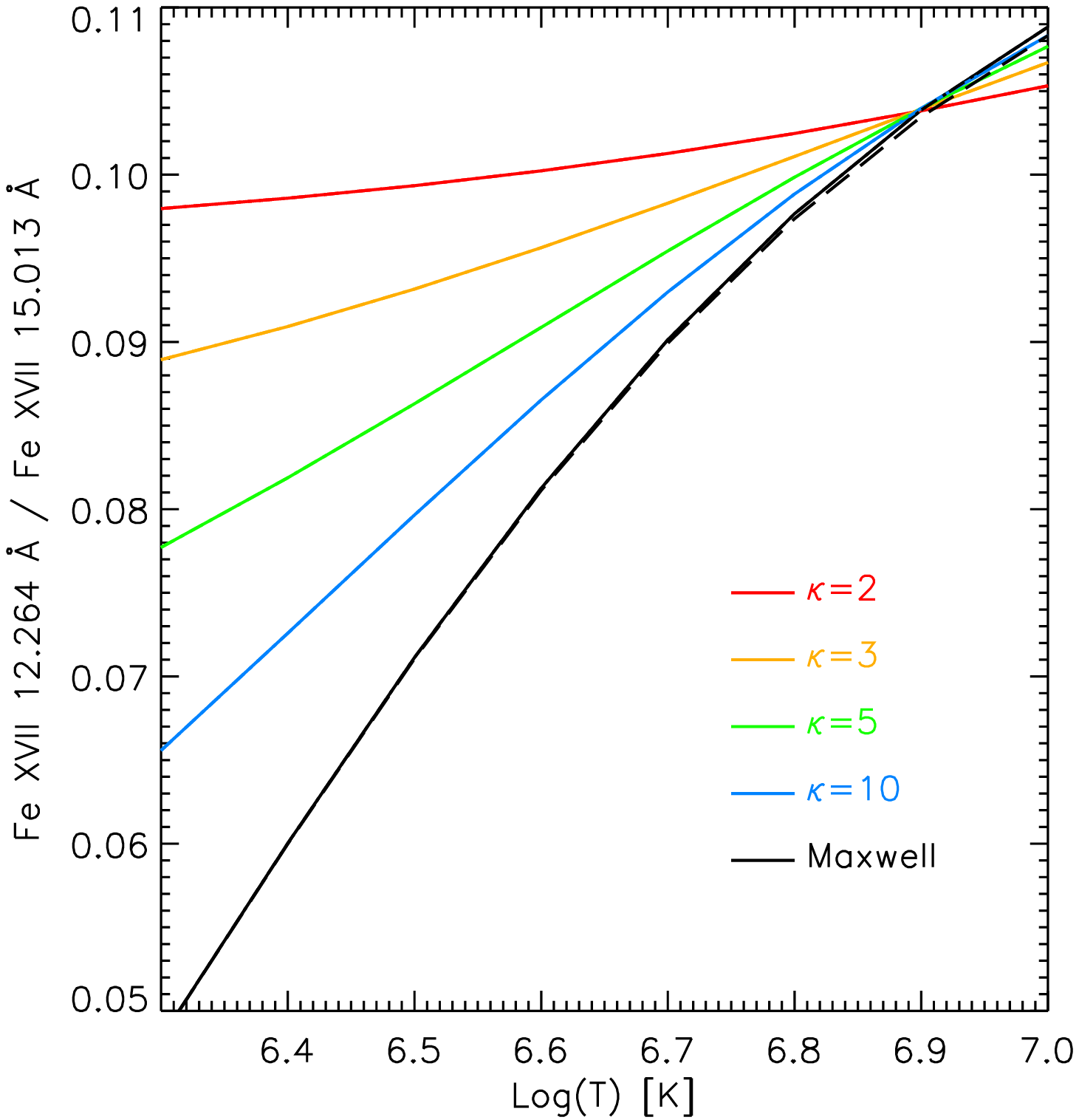}
   \caption{\textit{Left}: Effect of the electron distribution on the excitation rate of \ion{Fe}{XVII} line 15.01 \AA. \textit{Right}: Changes of the line intensity ratios with the electron distribution for the \ion{Fe}{XVII} 15.01 \AA\,/12.25 \AA\,. The line ratios are not density dependent.}
   \label{Fig:excit}
\end{figure*}
%
%
\section{The non-Maxwellian $\kappa$-distributions}
\label{Sect:2}

The $\kappa$-distribution (Fig. \ref{Fig:ioneq}, \textit{left}) is a distribution of electron velocities or energies, containing a characteristic power-law high-energy tail \citep{Owocki83,Livadiotis09}. In its isotropic energy form, the expression for a $\kappa$-distribution is (see, e.g., \citeauthor{Livadiotis15} \citeyear{Livadiotis15}, as well as \citeauthor{Livadiotis17} \citeyear{Livadiotis17}, chapter 4.3.1.3.2 therein)
\begin{equation}
	f(E,\kappa) \mathrm{d}E = A_{\kappa} \frac{2}{\sqrt{\pi} (k_\mathrm{B}T)^{3/2}} \frac{E^{1/2}\mathrm{d}E}{\left 			  (1+ \frac{E}{(\kappa - 3/2) k_\mathrm{B}T} \right)^{\kappa+1}}\,,
        \label{Eq:Kappa}
\end{equation}
where $E$ is electron kinetic energy and $k_\mathrm{B}$\,=\,1.38 $\times 10^{-16}$ erg\,K$^{-1}$ is the Boltzmann constant. The distribution is normalized to unity via
\begin{equation}
	A_{\kappa} = \frac{\Gamma(\kappa+1)}{\Gamma(\kappa-1/2) (\kappa-3/2)^{3/2}}\,.
	\label{Eq:A_kappa}
\end{equation}

The $\kappa$-distribution has two independent parameters (thermodynamic indices), $T$ and $\kappa$, see, for example, \citet{Livadiotis15,Livadiotis17}. The $T$ is the thermodynamic temperature \citep[see also][]{Livadiotis09}, coinciding with the definition in mean energy of the distribution $\left< E \right> = {3k_\mathrm{B}T}/{2}$. The parameter $\kappa$\,$\in$\,$\left(3/2,+\infty\right)$ describes the deviation from the Maxwellian. Maxwellian distribution is recovered for $\kappa$\,$\rightarrow$\,$\infty$, while $\kappa$\,$\rightarrow$\,3/2 represents the asymptotic, extremely non-Maxwellian situation.

We note that the $\kappa$-distribution can be thought of as consisting of a near-Maxwellian core and a high-energy power-law tail \citep{Oka13}. While the slope of the high-energy tail is given by $-(\kappa+1/2)$, the low-energy end of the distribution can be well approximated by a Maxwellian with a temperature $T_\mathrm{M}$\,=\,$T(\kappa-3/2) / (\kappa +1)$ \citep{Meyer-Vernet95,Livadiotis09}. This permitted \citet{Oka13} to calculate the relative number of the particles in the high-energy tail, as well as the energy carried by these particles. For example, for a moderately non-Maxwellian situation of $\kappa$\,=\,5, the high-energy tail contains $\approx$16\% of particles carrying $\approx$42\% of energy. For a more extreme $\kappa$\,=\,2 situation, the high-energy tail contains $\approx$36\% particles carrying more than 83\% of total kinetic energy of the distribution \citep[Fig. 1b of][]{Oka13}.

%
%
\section{Spectral synthesis}
\label{Sect:3}
\subsection{Spectral line emissivities}
\label{Sect:3.1}

The spectral synthesis of optically thin \ion{Fe}{XVII} and \ion{Fe}{XVIII} lines here is done analogously to \citet[][hereafter, Paper\,I]{Dudik14b}. The intensity $I_{ji}$ of a spectral line corresponding to a transition $j \to i$ between levels $j > i$ is given by \citep[cf.,][]{Mason94,Phillips08}
\begin{equation}
	I_{ji} = \int A_X G_{X,ji}(T,N_\mathrm{e},\kappa) N_\mathrm{e} N_\mathrm{H} \mathrm{d}l\,,
	\label{Eq:line_intensity}
\end{equation}
where $l$ is the line of sight, $N_\mathrm{e}$ and $N_\mathrm{H}$ are the electron and hydrogen densities, respectively, $A_X$ is the abundance of element $X$, and $G_{X,ji}(T,N_\mathrm{e},\kappa)$ is the line contribution function, given by the expression
\begin{equation}
	G_{X,ji}(T,N_\mathrm{e},\kappa) = \frac{hc}{\lambda_{ji}} \frac{A_{ji}}{N_\mathrm{e}} \frac{N(X_j^{+k})}{N(X^{+k})} \frac{N(X^{+k})}{N(X)}\,.
	\label{Eq:G(T)}
\end{equation}
There, $\lambda_{ji}$ is the line wavelength, $hc/\lambda_{ji}$ is the corresponding photon energy, $A_{ji}$ is the Einstein coefficient for the spontaneous emission, $N(X_j^{+k}) / N(X^{+k})$ is the fraction of the ion $X^{+k}$ with the electron on the upper excited level $j$, and $N(X^{+k})/N(X)$ is the relative ion abundance of the ion $X^{+k}$.

\subsection{Ionization and excitation equilibria for the $\kappa$-distributions}
\label{Sect:3.2}

Both the relative ion abundance $N(X^{+k})/N(X)$ and the excitation fraction $N(X_j^{+k}) / N(X^{+k})$ are functions of $\kappa$. This comes from the fact that the individual ionization, recombination, excitation, and deexcitation rates all depend on $\kappa$ \citep[e.g.,][and references therein]{Dzifcakova13,Dudik14b,Dzifcakova15}. These ratios are calculated by assuming ionization and excitation equilibrium, respectively, in other words, these fractions are assumed to be time-independent. For the excitation equilibrium, this assumption is justified since the equilibration times are small, depending on the corresponding $A_{ji}$ values and the excitation rates $C_{ij}^\mathrm{e}$. At densities of log$(N_\mathrm{e}$\,[cm$^{-3}$])\,=\,10, the $N_\mathrm{e}C_{ij}^\mathrm{e}$ is of the order of 0.1--1\,s$^{-1}$ for allowed lines of \ion{Fe}{XVII} and \ion{Fe}{XVIII} at log$(T$\,[K])\,=\,6.6  (e.g., Fig. \ref{Fig:excit}), that is, temperatures typical of an active region cores.

The ionization equilibration timescales can however be larger. For iron, they are of the order of ten seconds at log$(T$\,[K])\,=\,6.6, according to Fig.\,1 of \citet{Smith10}. These results were calculated for the Maxwellian distribution; however, the ionization equilibration timescales will not be significantly different for the $\kappa$-distributions at these temperatures, since the ionization and recombination rates for $\kappa$-distributions do not differ by more than about a half an order of magnitude at log($T$\,[K])\,=\,6.6 \citep[see Fig. 2 of][]{Dzifcakova13}, with the ionization rates increasing for lower $\kappa$ compared to a Maxwellian.

Therefore, the results obtained in this paper will be valid for active region cores that do not display significant temporal variability on the order of ten seconds or less. 

Finally, the relative ion abundance of \ion{Fe}{XVII} and \ion{Fe}{XVIII} obtained by \citet{Dzifcakova13} are shown in Fig. \ref{Fig:ioneq}. There, the dashed and full lines stand for \ion{Fe}{XVII} and \ion{Fe}{XVIII}, respectively, while the different colors denote different values of $\kappa$. The typical widening of the peaks of the relative ion abundances for lower $\kappa$ is evident. Furthermore, for $\kappa$\,$\lesssim$\,2, a shift of the peaks to higher $T$ occurs. This shift is only about 0.05 in log($T$\,[K]) for both \ion{Fe}{XVII} and \ion{Fe}{XVIII} (Fig. \ref{Fig:ioneq}). The atomic data behind the individual ionization and recombination rates correspond to CHIANTI database, versions 6--8 \citep{Dere97,Dere07,Dere09,Landi13,DelZanna15a}.

The collisional excitation and de-excitation rates are calculated by direct integration of the original collision strengths $\Omega_{ji}$ (non-dimensionalized excitation cross-sections) over the $\kappa$-distributions using the method of \citet{Bryans06}, described in detail in Paper\,I (see Eqs. (4)--(16) therein). 

The collision strengths $\Omega_{ji}$ as well as the $A_{ji}$ values were calculated by \citet{Liang10b} for \ion{Fe}{XVII} and \citet{DelZanna06} for \ion{Fe}{XVIII}. We note that these atomic data are also consistent with CHIANTI, version 8.0.7 \citep{Dere97,DelZanna15a}.

\subsection{Corrections to level populations due to ionization and recombination}
\label{Sect:3.3}
The line emissivity calculations outlined in previous sections are done in the coronal approximation. This means  that the populations of the excited levels of the \ion{Fe}{XVII} and \ion{Fe}{XVIII} are negligible compared to the population of the ground level. This means that the ionization and recombination happens dominantly from and to the ground level, respectively. However, this approximation does not always hold for \ion{Fe}{XVII} \citep{Doron02} as well as for higher ionization states of Fe \citep{Gu03}.

\citet{Doron02} calculated the \ion{Fe}{XVII} emissivities by including the neighboring ionization stages of \ion{Fe}{XVI} and \ion{Fe}{XVIII}, while assuming that their relative ion abundances are imposed. These authors showed that some strong lines of \ion{Fe}{XVII}, such as the 2$p$--3$s$ at 16.78, 17.05, and 17.10\,\AA~(see Appendix \ref{Appendix:Transitions}) contain significant contributions, exceeding 50\% in some cases, from resonant excitation and dielectronic recombination. The resonant excitation is dominant at coronal temperatures, while the importance of dielectronic recombination increases with increasing temperatures, when the relative abundance of \ion{Fe}{XVII} is small compared to \ion{Fe}{XVIII} (see also Fig. \ref{Fig:ioneq}). Contrary to that, the contribution of the inner-shell ionization changes the level populations by less than about 3\%. \citet{Gu03} extended these results to \ion{Fe}{XVII}--\ion{Fe}{XXIV}, while providing tabulated values of the  rates of individual processes. 

In the CHIANTI atomic database and software, the contributions of resonant excitation for \ion{Fe}{XVII} and \ion{Fe}{XVIII} is already included in the electron excitation rates, since these are based on the collision strengths $\Omega_{ji}$ which themselves contain the resonances. The contributions to the level populations from level-resolved dielectronic recombination and collisional ionization are included via the correction factors \citep[see Eq. (10) of][]{Landi06} that are based on the ionization and recombination rates of \citet{Gu03} and are applied to the level populations calculated in coronal approximation.

We note that \citet{Gu03} calculated the level-resolved ionization and recombination rates for the Maxwellian distribution. Furthermore, these rates are effective since they include the contribution of cascading. For this reason, the corresponding non-Maxwellian rates cannot be easily calculated using the method of \citet{Dzifcakova13}. This is because the contribution of cascading cannot be separated from the ionization and recombination to upper levels. In addition, \citet{Gu03} included many upper levels (up to $n$\,=\,45 compared to only levels up to $n$\,=\,6 in CHIANTI). Therefore, including level-resolved ionization and recombination to $n$\,$\leq$\,6 levels for non-Maxwellians would still not contain the contributions from cascading from higher levels. Work is however ongoing in this respect (Dzif\v{c}\'{a}kov\'{a} et al., in preparation).

For the above reason, we calculate the level populations for non-Maxwellians \textit{without} the correction factors. For the Maxwellian distribution, the corresponding error created by not including the correction factors is always less than about 8\% in \ion{Fe}{XVII} X-ray line intensities at temperatures log($T$\,[K])\,$\leq$\,6.6, with only some lines being affected \citep[see, e.g.,][]{Doron02,Gu03}. For the \ion{Fe}{XVIII}, the X-ray line intensities change by $\lesssim$\,9\%, with the lines 15.83--16.07\,\AA~being the most affected. Notably, the 93.93\,\AA~line is affected by only about 2\% at these temperatures. In this respect, at coronal temperatures, the corrections to level populations due to ionization and recombination are secondary effects, as also noted by \citet{DelZanna11b}. Nevertheless, to indicate the influence of these corrections for the Maxwellian distribution, the corresponding ratios of line intensities calculated with the correction factors are shown by dashed black lines in Figures \ref{Fig:Diag_single-ion_fe17}--\ref{Fig:Diag_magixs_aia}. We note that the contributions from dielectronic recombination increase with increasing temperature. Therefore, some ratios can become less reliable for diagnostics of non-Maxwellian distributions in case of high-temperature plasma, for example in flares. Details for respective cases are provided in Sect. \ref{Sect:4.2}.

%
\begin{figure*}
   \centering
   \includegraphics[width=18.6cm,viewport=20 0 850 335,clip]{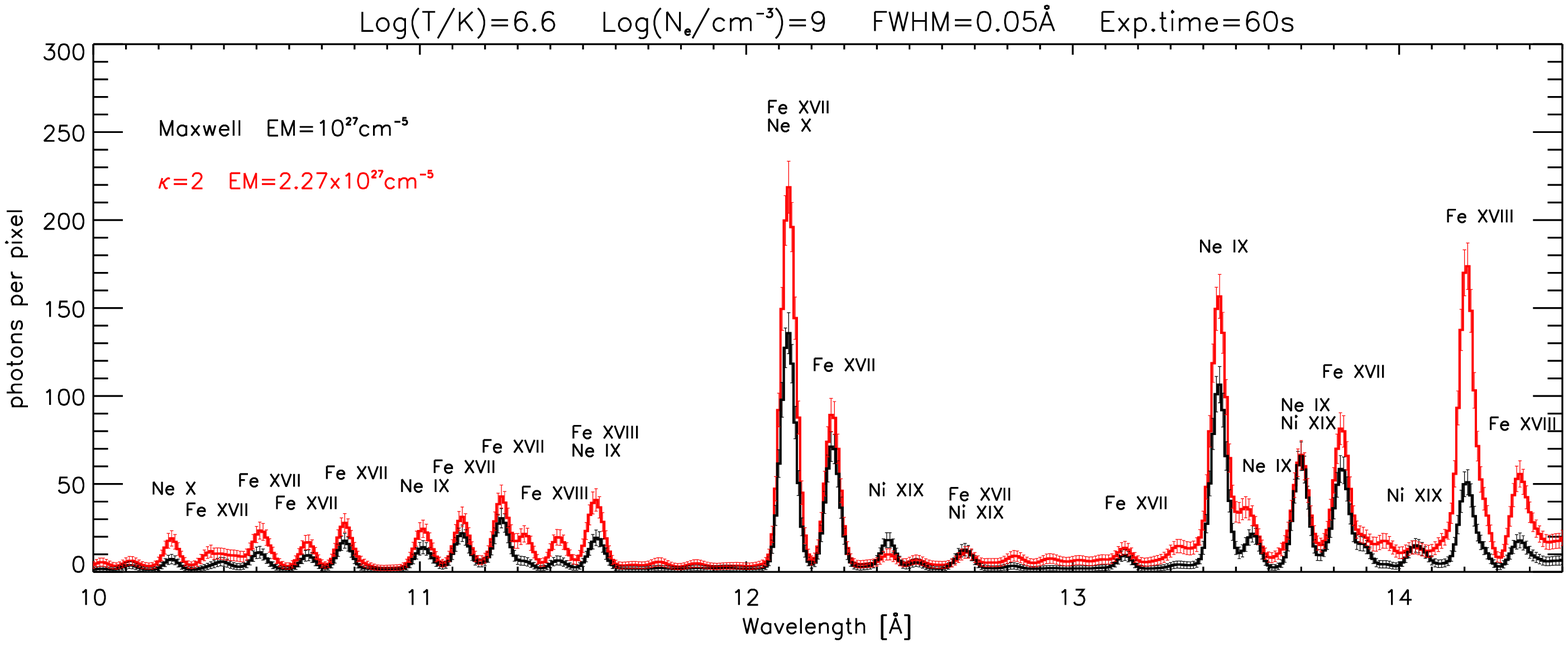}
   \includegraphics[width=18.6cm,viewport=20 0 850 335,clip]{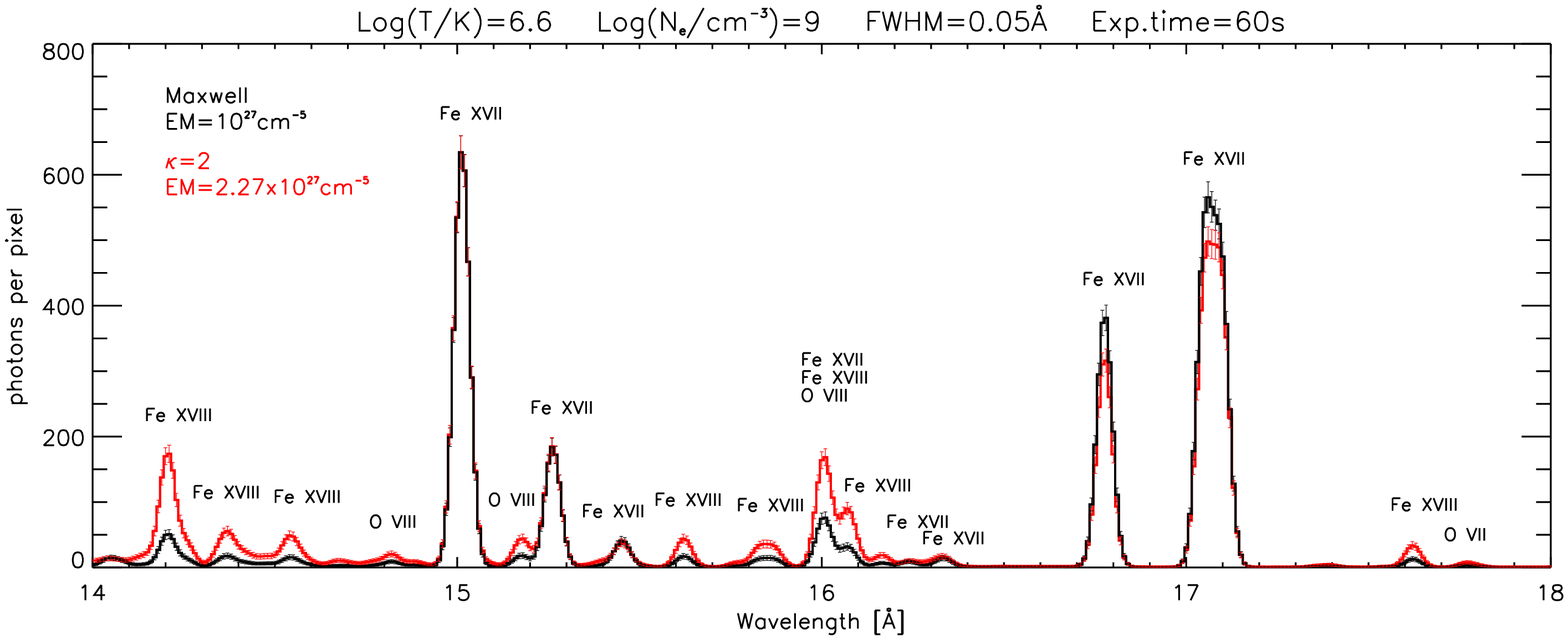}
   \caption{Example synthetic \textit{MaGIXS} spectra at 10--18\,\AA. The Maxwellian spectrum (\textit{black}) is compared here with a spectrum obtained for $\kappa$=2 (\textit{red}). Both spectra were calculated for log($T$\,[K])\,=\,6.6 and $N_\mathrm{e}$\,=\,10$^9$\,cm$^{-3}$. Error-bars indicate photon noise uncertainty in a 60\,s integration time. The assumed FWHM of the lines is 0.05\,\AA. Principal ions forming a particular spectral line are indicated. For details on blends see Appendix \ref{Appendix:Transitions}.}
   \label{Fig:spec}%
\end{figure*}
%
%

%
\begin{figure*}
	\centering
	\includegraphics[width=8.8cm]{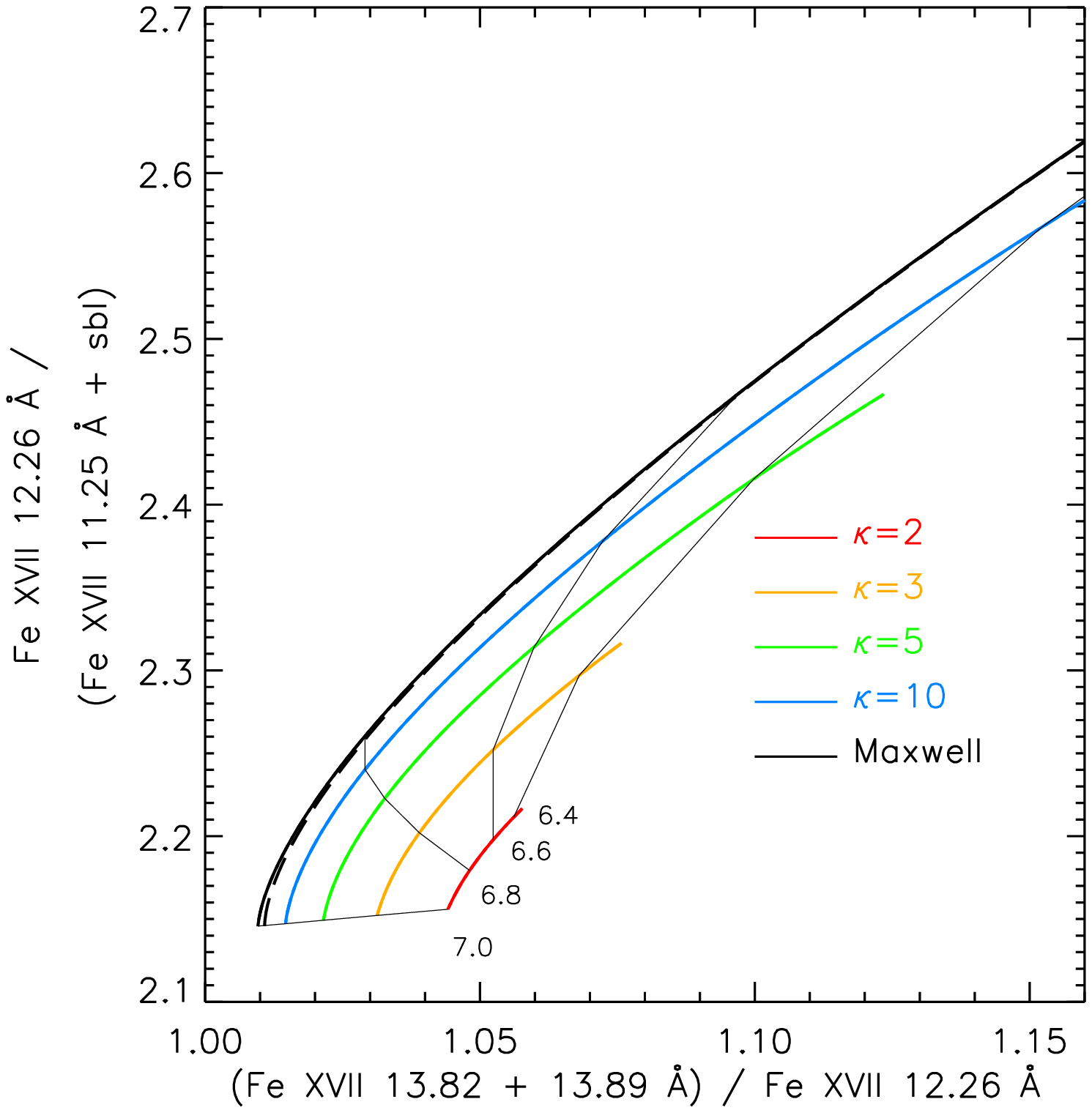}
	\includegraphics[width=8.8cm]{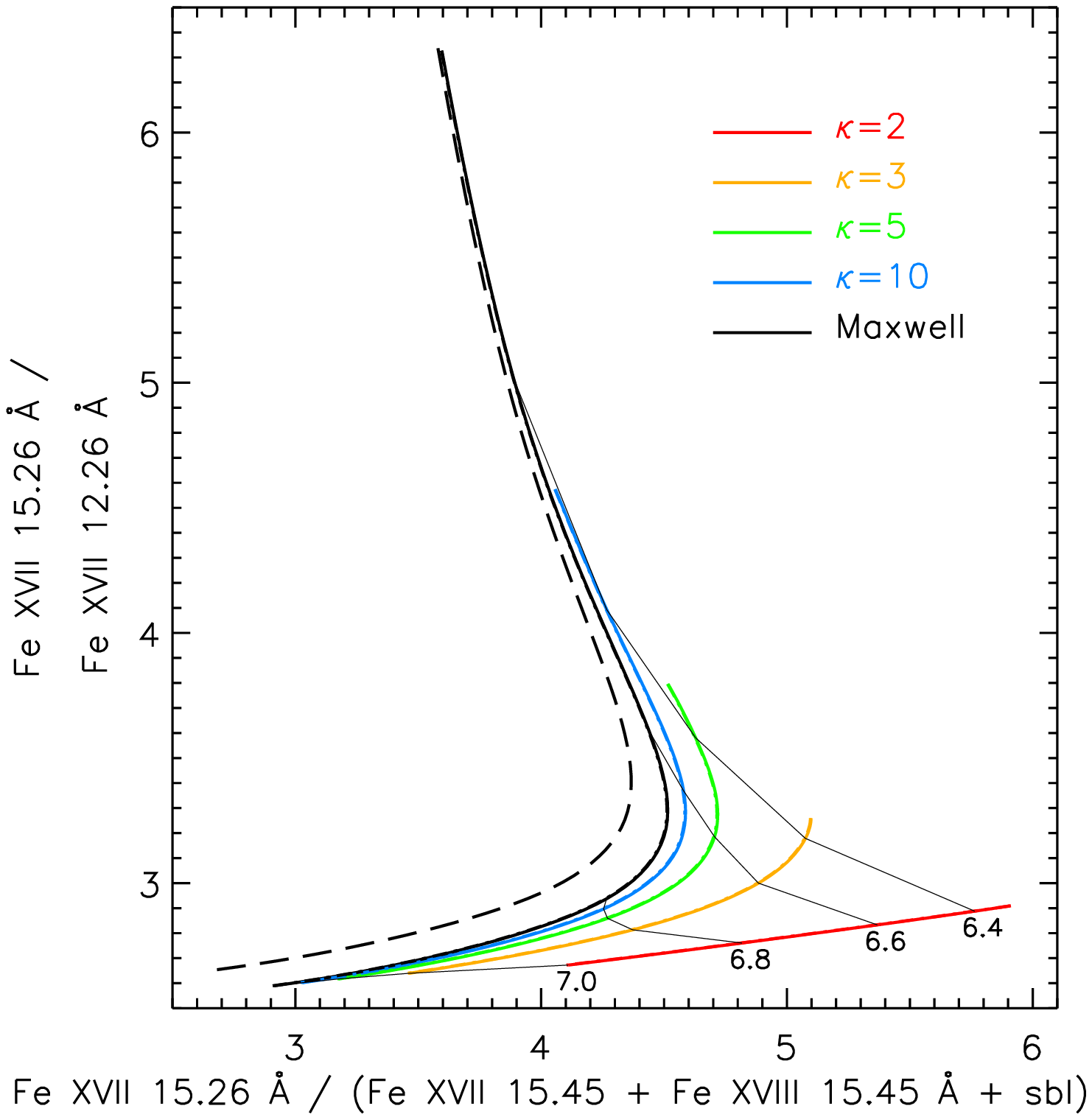}

	\caption{Diagnostics of $\kappa$ and temperature from \ion{Fe}{XVII} from the \textit{MaGIXS} spectra. Ratios of line intensities (in photon units) are indicated, with wavelengths of the lines involved listed together with the main self-blends. Individual colors denote the distribution, with black for Maxwellian, and blue, green, orange, and red for $\kappa$-distribution with $\kappa$\,=\,10, 5, 3, and 2, respectively. Full lines correspond to spectra with for $N_\mathrm{e}\,=\,$10$^9$\,cm$^{-3}$, while the dot-dashed lines correspond to $N_\mathrm{e}\,=\,$10$^{11}$\,cm$^{-3}$. Black long-dashed lines (available only for the Maxwellian distribution) show the effects of the recombination and ionization to and from the upper excited levels, together with cascades. See Sect. \ref{Sect:3.3} for more details.}
\label{Fig:Diag_single-ion_fe17}
\end{figure*}
%

%
\begin{figure*}
	\centering
	\includegraphics[width=8.8cm]{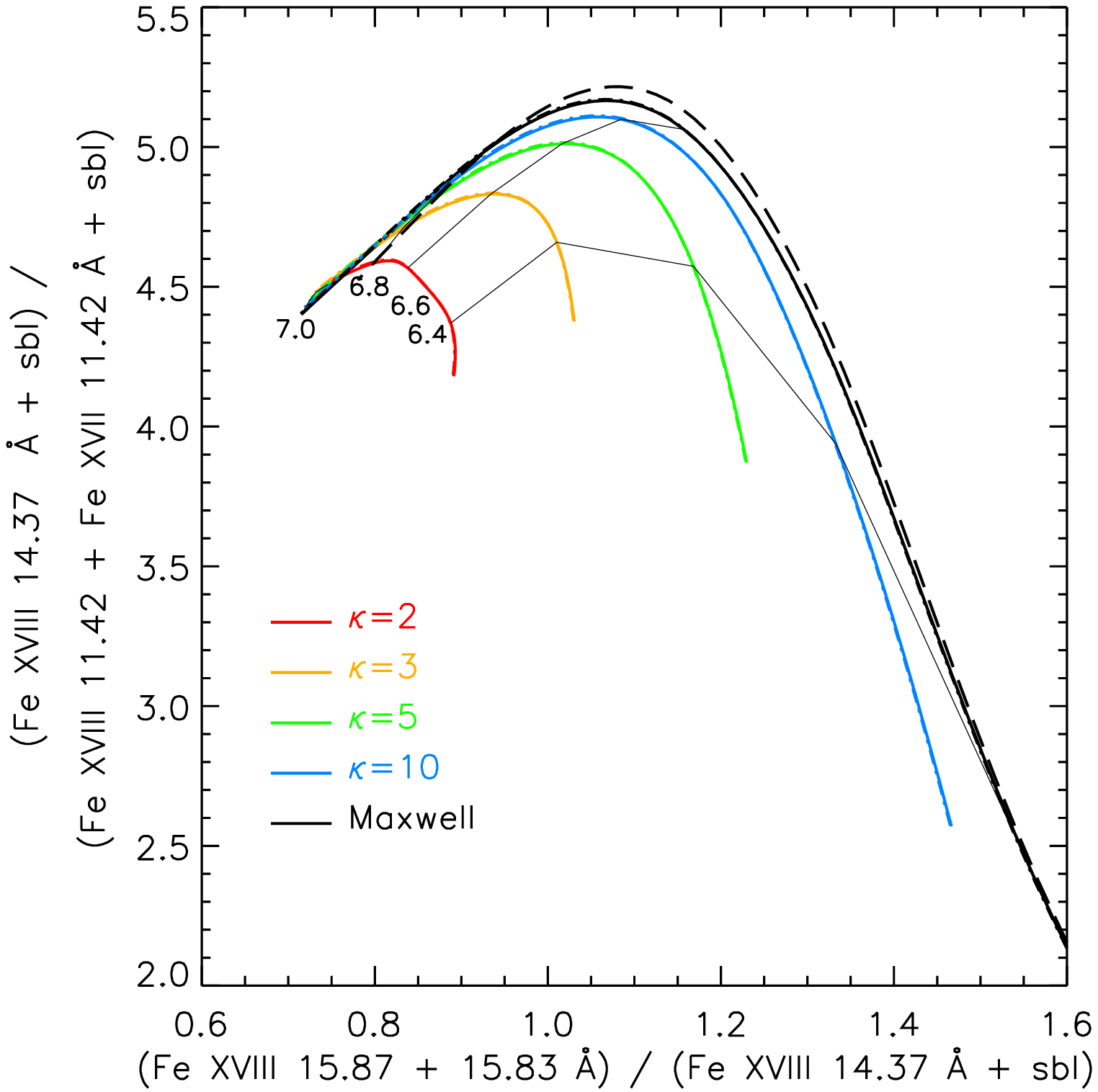}
	\includegraphics[width=8.8cm]{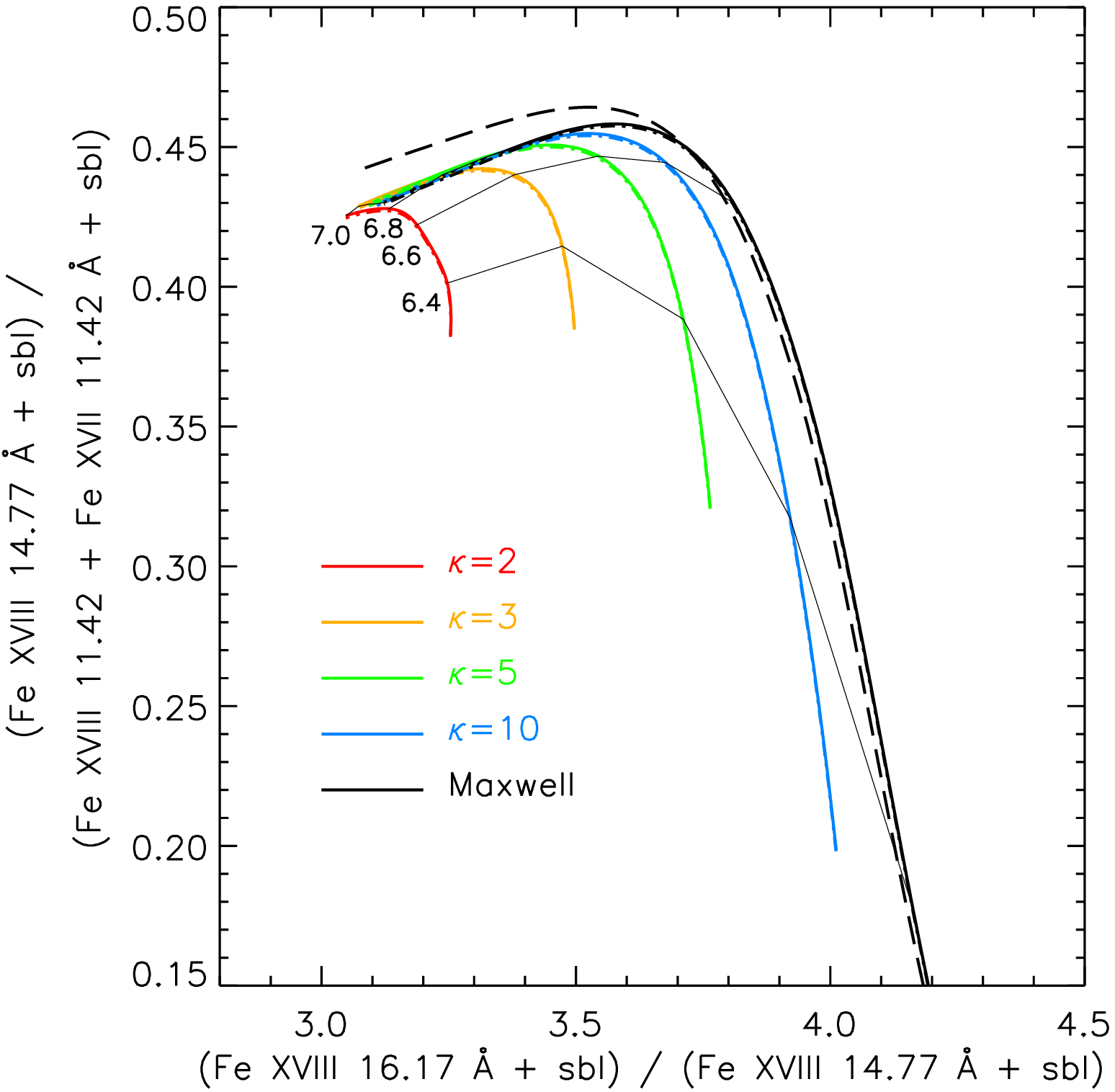}
	\caption{Same as in Fig. \ref{Fig:Diag_single-ion_fe17}, but for \ion{Fe}{XVIII}.}
\label{Fig:Diag_single-ion_fe18}
\end{figure*}
%

%
\begin{figure*}
	\centering
	\includegraphics[width=8.8cm]{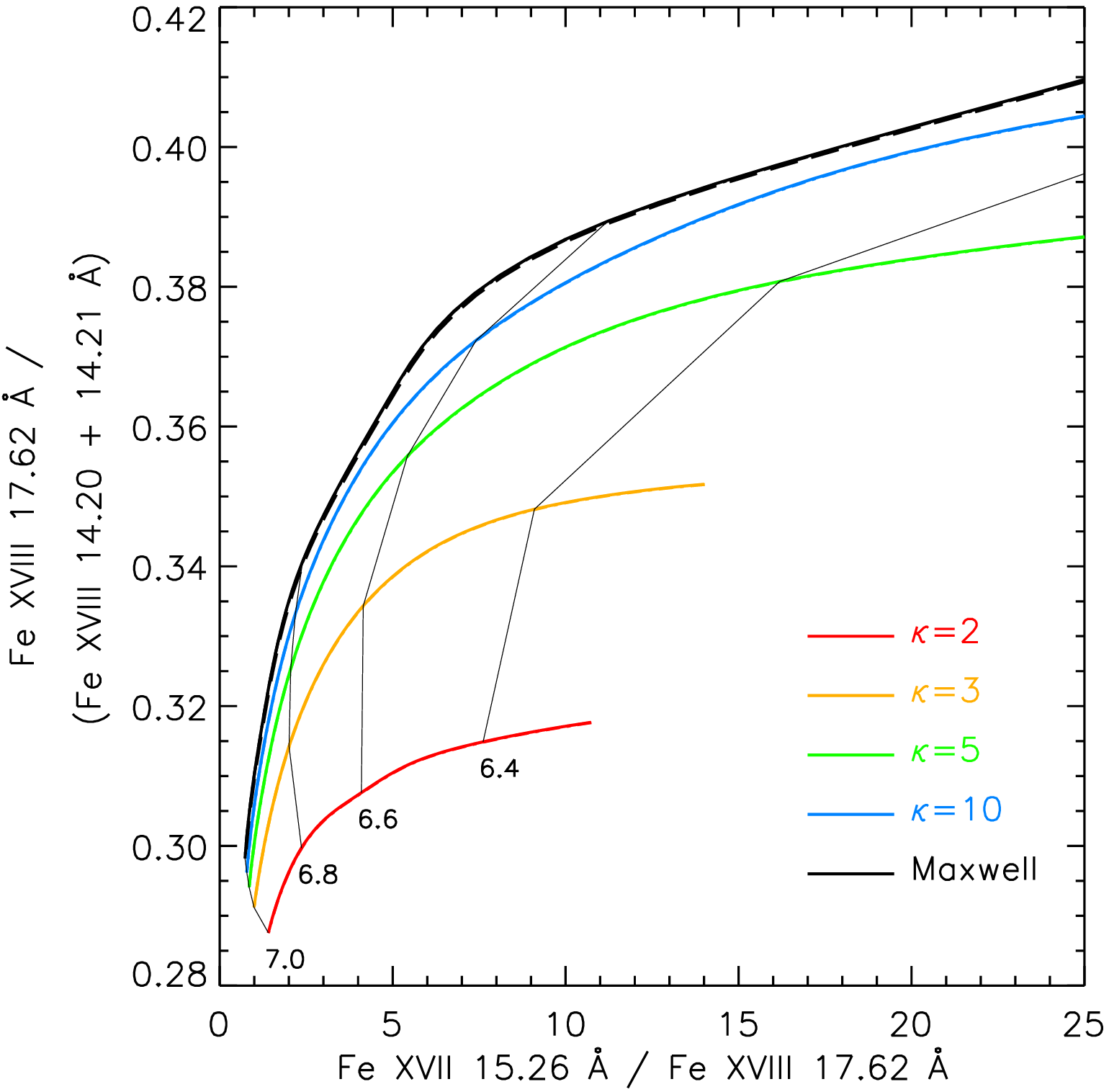}
	\includegraphics[width=8.8cm]{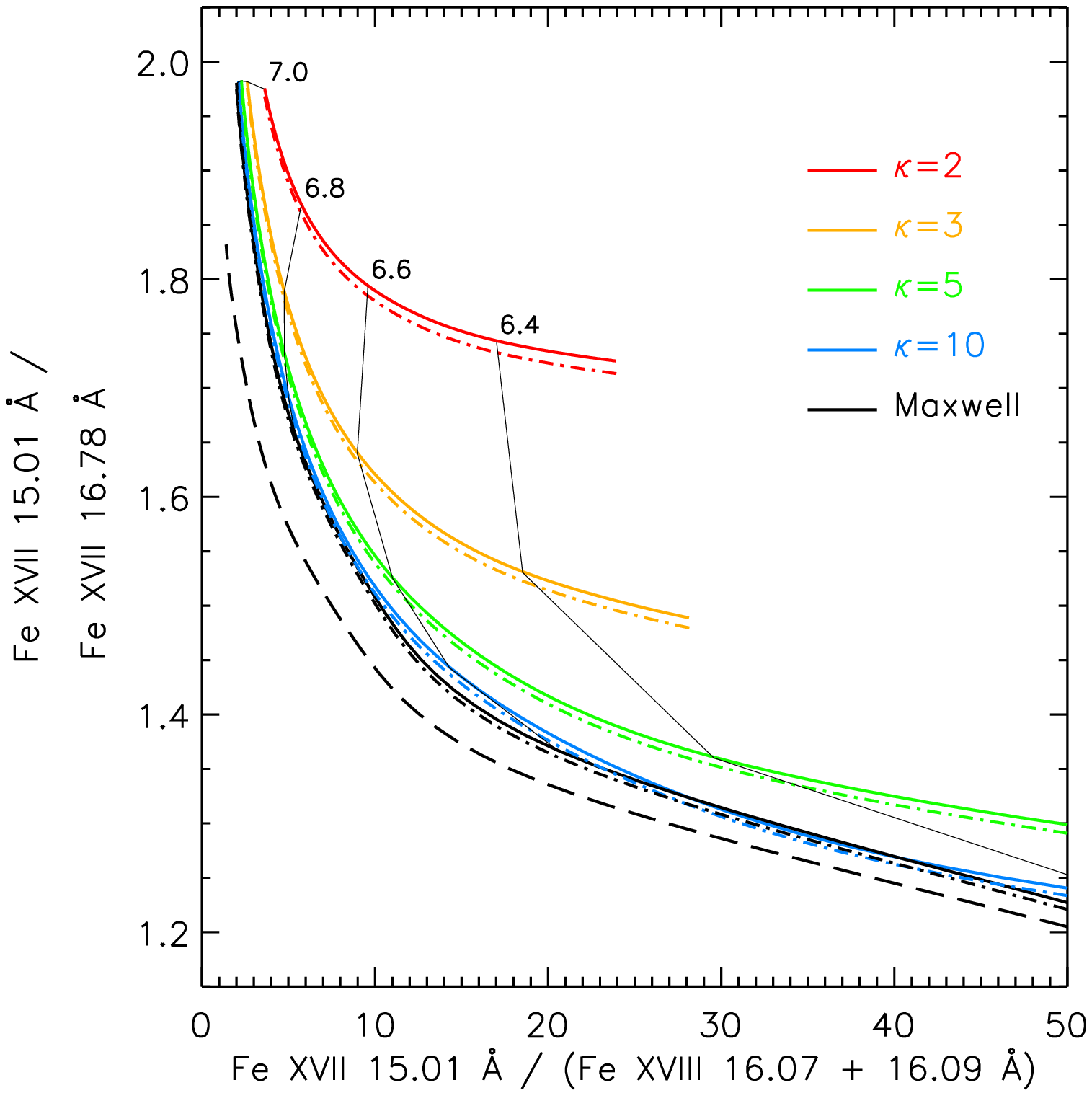}
	\includegraphics[width=8.8cm]{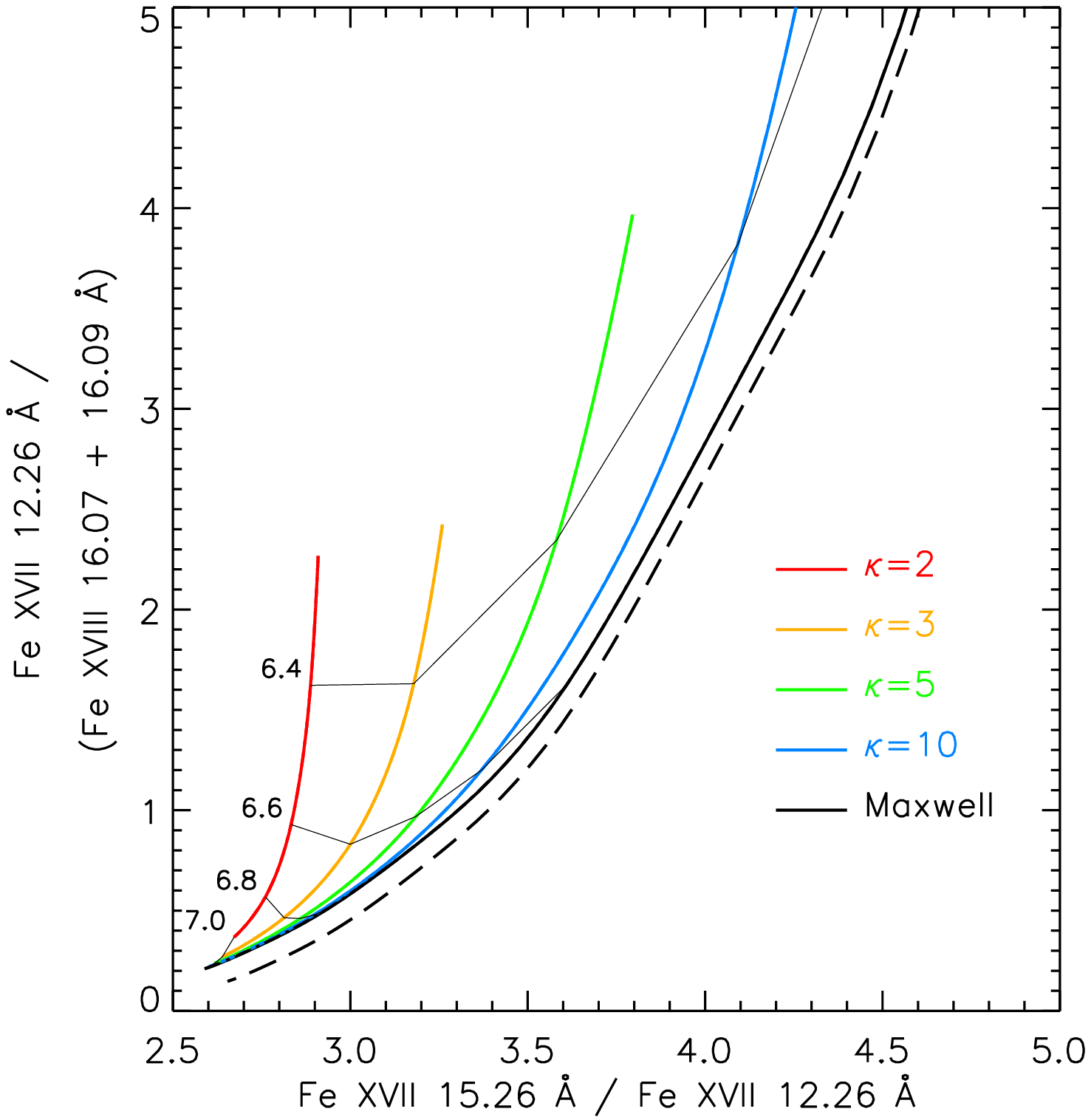}
	\includegraphics[width=8.8cm]{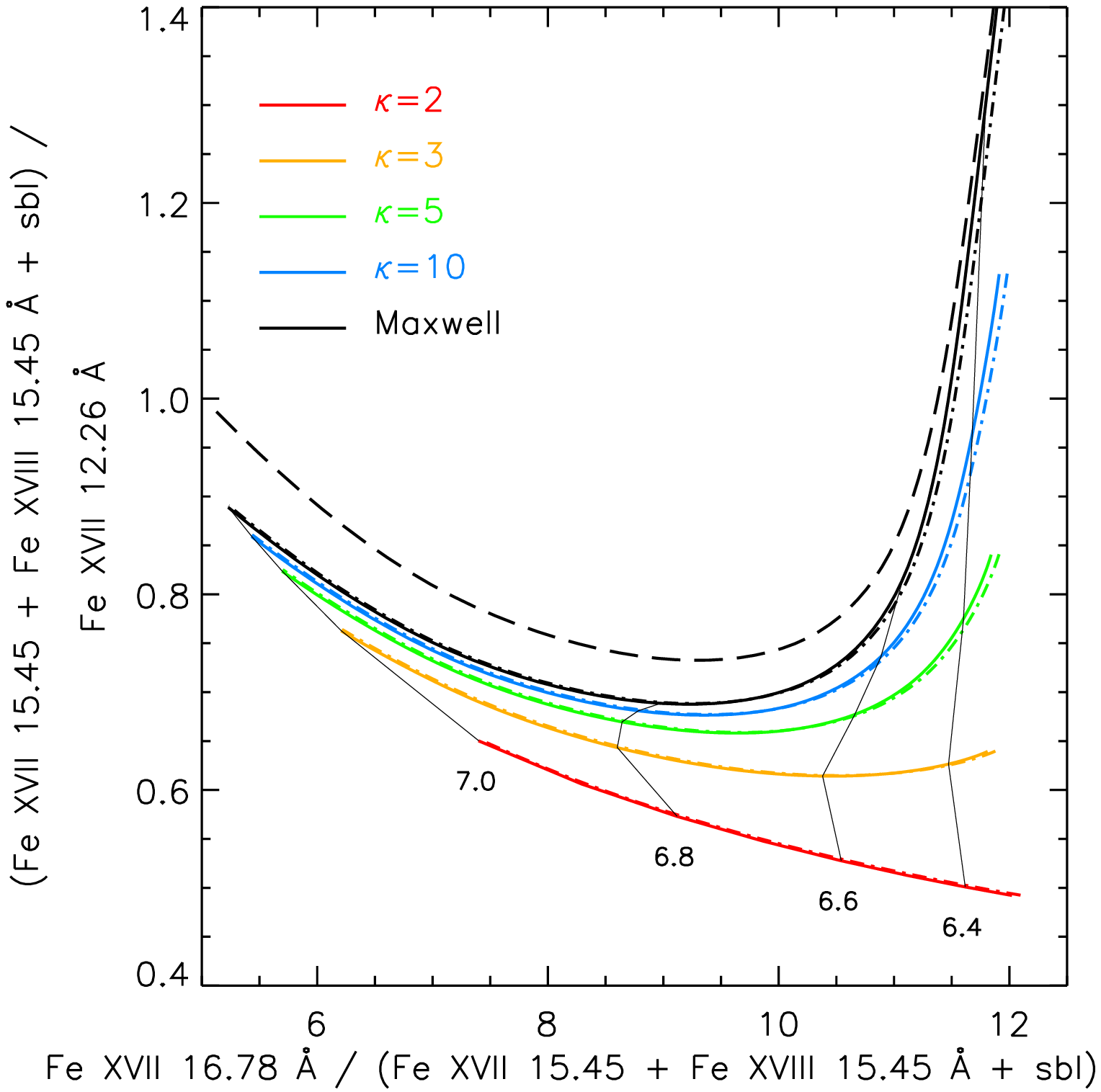}
	\caption{Same as Fig. \ref{Fig:Diag_single-ion_fe17}, but for diagnostics involving both \ion{Fe}{XVII} and \ion{Fe}{XVIII}.}
\label{Fig:Diag_multi-ion}
\end{figure*}
%

%
\begin{figure*}
	\centering
	\includegraphics[width=8.8cm]{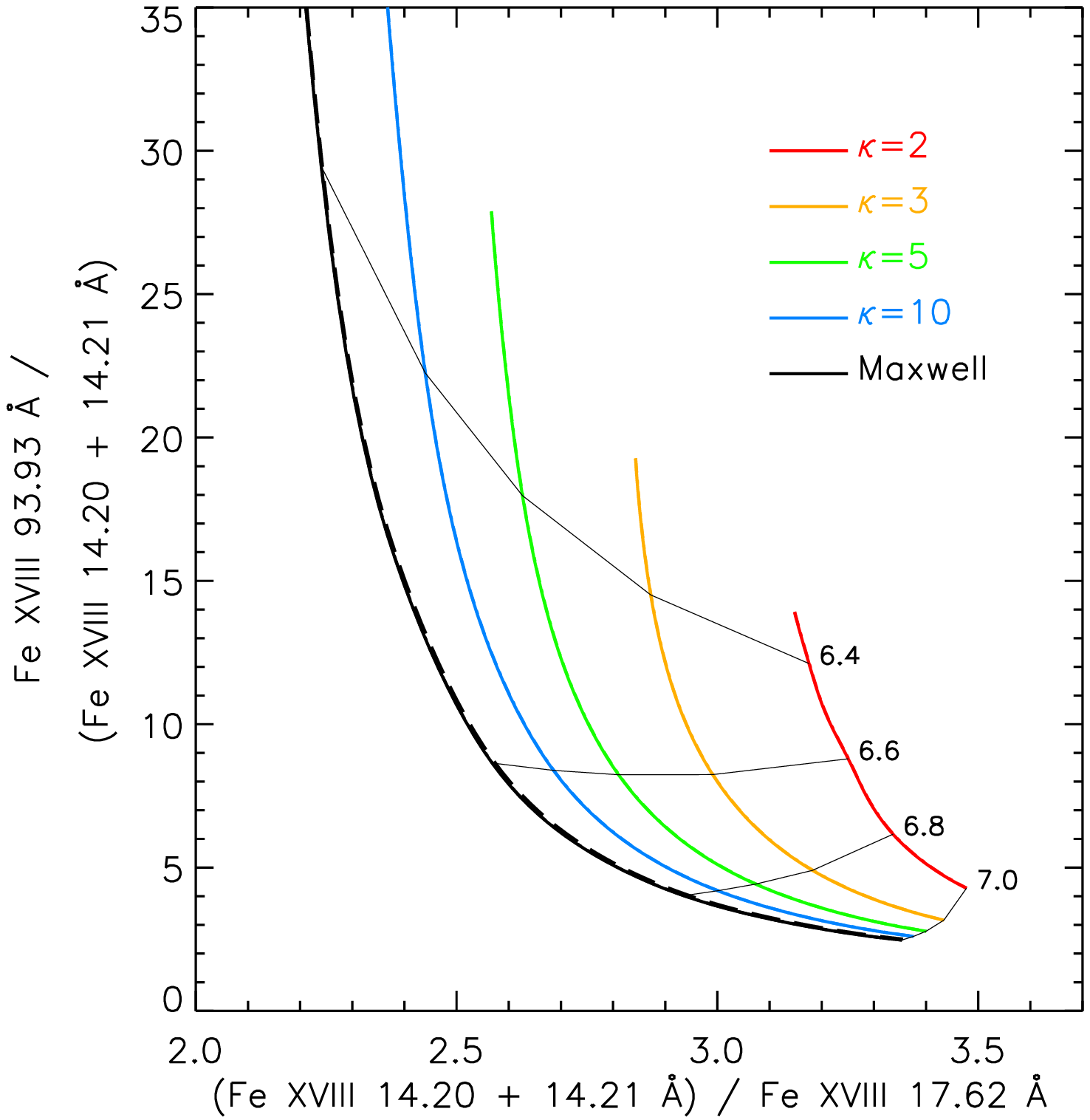}
	\includegraphics[width=8.8cm]{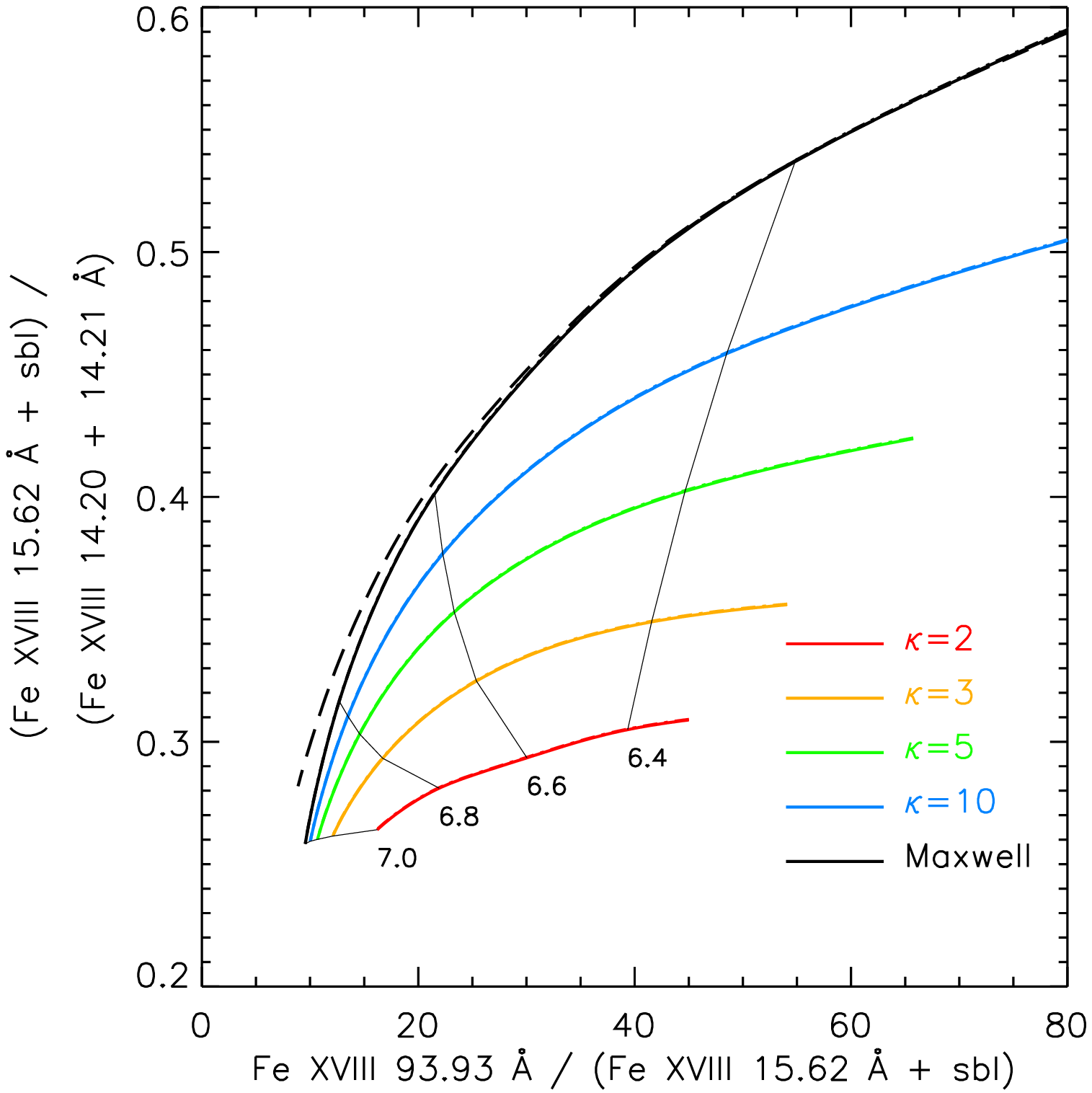}
	\includegraphics[width=8.8cm]{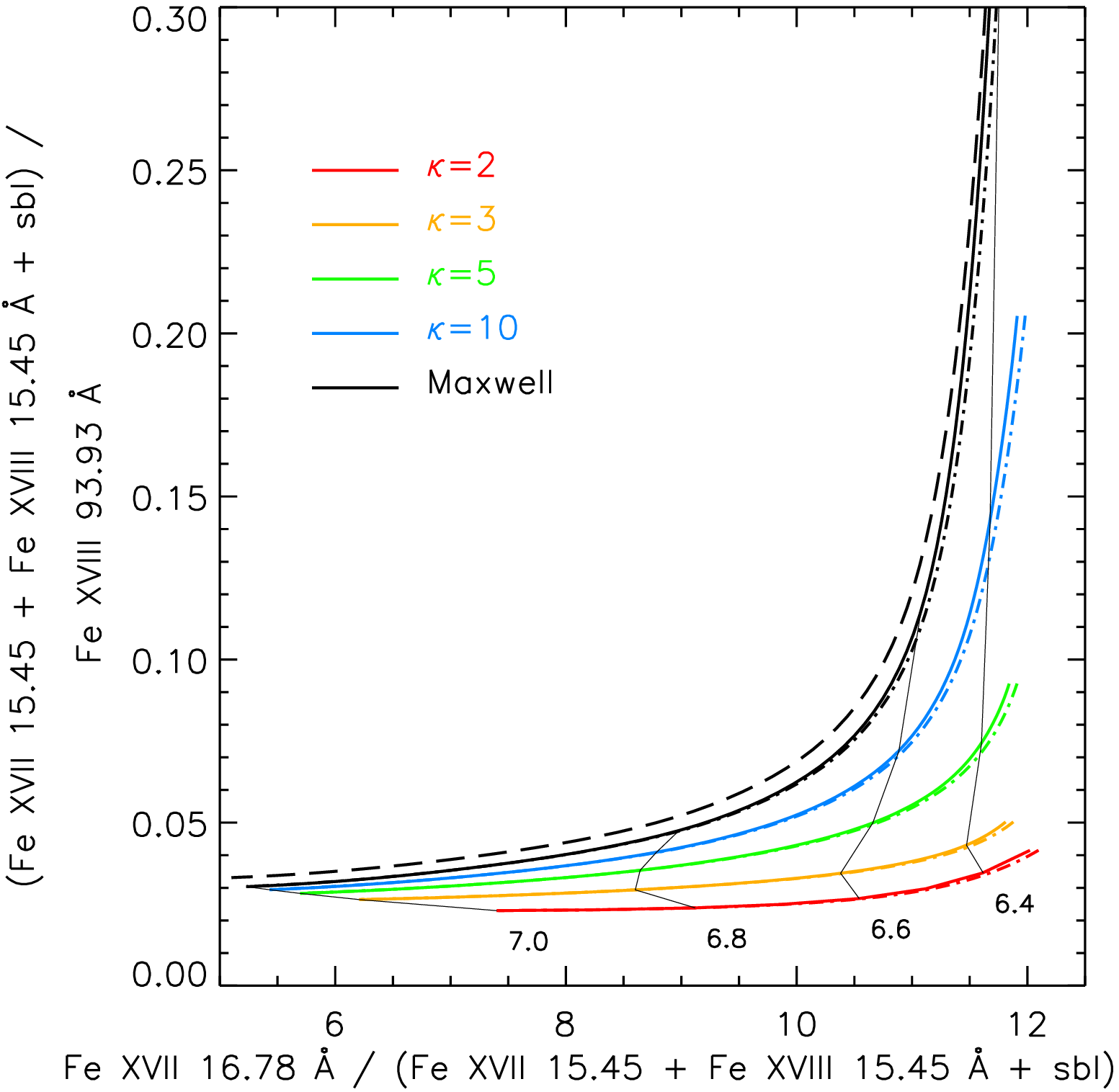}
	\includegraphics[width=8.8cm]{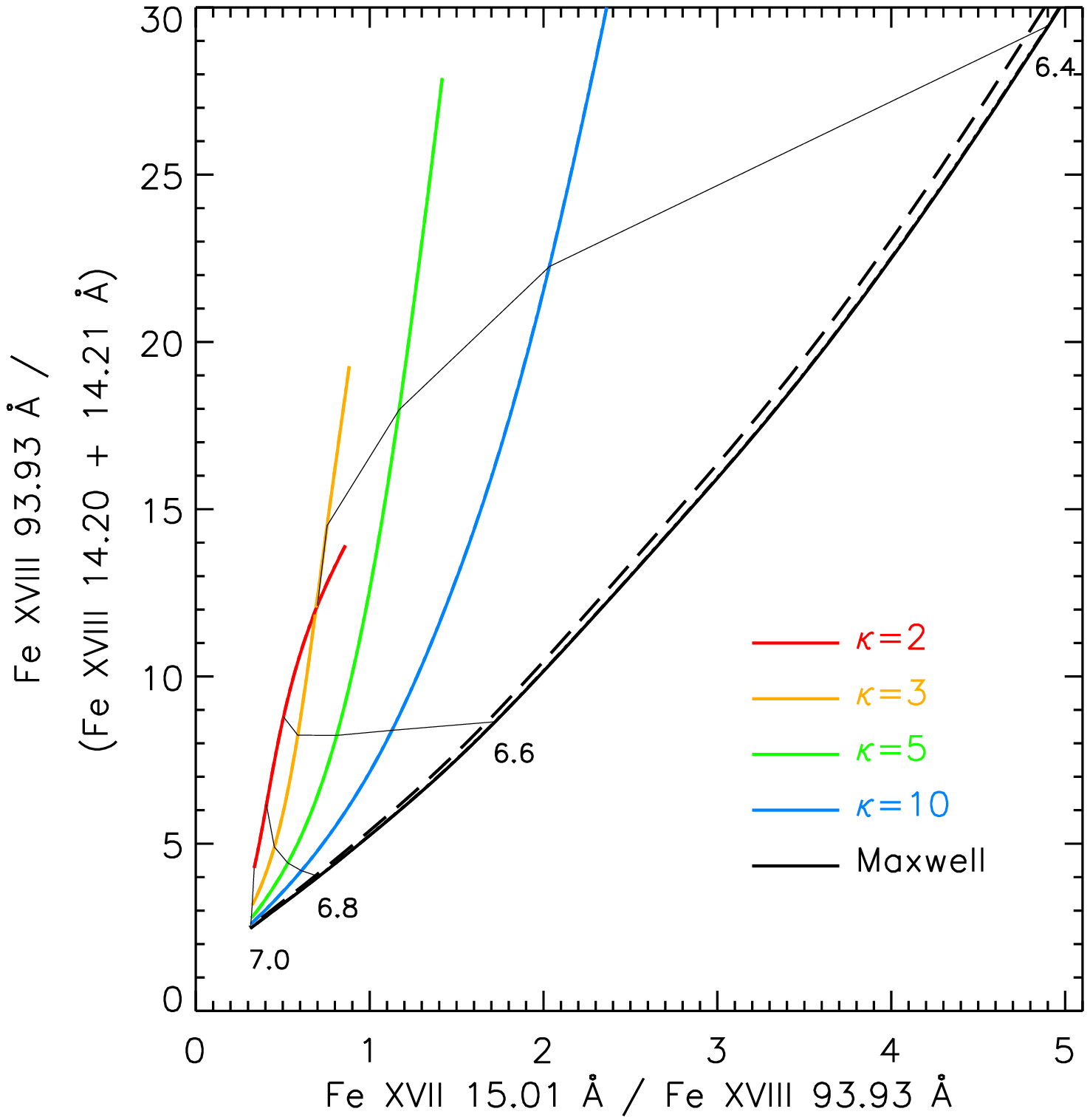}
\caption{Same as Fig. \ref{Fig:Diag_single-ion_fe17}, but for diagnostics using the \textit{MaGIXS} X-ray lines in combination with the \ion{Fe}{XVIII} 93.93\,\AA~line observed by \textit{SDO}/AIA.}
\label{Fig:Diag_magixs_aia}
\end{figure*}
%

%
%
%
%
%
\section{Results}
\label{Sect:4}

\subsection{Synthetic \textit{MaGIXS} line spectra}
\label{Sect:4.1}

We calculated the synthetic spectra of \ion{Fe}{XVII}--\ion{Fe}{XVIII} using the method outlined in Sect. \ref{Sect:3}. In addition, lines of other ions and elements (such as \ion{Ne}{IX}, \ion{Ni}{XIX}, and \ion{O}{VIII}) producing lines in the \textit{MaGIXS} spectral window were calculated using the KAPPA database \citep{Dzifcakova15}. The synthetic \textit{MaGIXS} spectra are shown in Fig. \ref{Fig:spec}. These are example spectra, calculated in isothermal conditions for log($T$\,[K])\,=\,6.6, assuming the Maxwellian (black) as well as the $\kappa$-distribution with $\kappa$\,=\,2 (red color). The spectra were calculated in the range of 10--18\,\AA, where many prominent \ion{Fe}{XVII} and \ion{Fe}{XVIII} lines are present (see also Appendix \ref{Appendix:Transitions}). The line intensities were calculated in photon units (phot\,cm$^{-2}$\,s$^{-1}$\,sr$^{-1}$), then converted to counts based on the \textit{MaGIXS} effective area as a function of wavelength, assuming pixel size and slit width of 2$\farcs$5, integration time of 60\,s, and emission measure $EM$\,=\,10$^{27}$\,cm$^{-5}$ typical of small active regions \citep[][Figs. 1 and 6 therein]{Warren12}. Finally, the spectra were convolved with Gaussian line profiles using a constant FWHM of 0.05\,\AA~(with pixel size of 0.01\,\AA). Photon noise uncertainties are indicated by error-bars. The read noise was not added as it is expected to be significantly smaller than the photon noise. Finally the emission measure for the spectrum with $\kappa$\,=\,2 has been multiplied by a factor of 2.27 so that the line intensity ratio of the \ion{Fe}{XVII} 15.01\,\AA\,/\,15.26\,\AA~lines is the same for both Maxwellian and a $\kappa$-distribution. In this way, the changes in line intensities for $\kappa$\,=\,2 with respect to the Maxwellian should become readily apparent. 

From the synthetic spectra shown in Fig. \ref{Fig:spec} it is apparent that some lines are quite sensitive to the $\kappa$-distributions. The strong \ion{Fe}{XVII} 2p$^6$\,--\,2p$^5$\,3s lines at 16.78\,\AA, and the 17.05 and 17.09\,\AA~selfblend are decreased for the $\kappa$\,=\,2. Largest differences however occur for weaker lines of both \ion{Fe}{XVII} and \ion{Fe}{XVIII}, many of which however have less than about 5\% intensity relative to \ion{Fe}{XVII} 15.01\,\AA. Detailed diagnostics using these lines are given in Sect. \ref{Sect:4.2}. To further show the changes with $\kappa$ in these weaker lines, as well as their photon noise uncertainties, the spectrum of Fig. \ref{Fig:spec} is also plotted in Appendix \ref{Appendix:Spectrum_log} using a logarithmic intensity scale. 

Other conspicuous examples of lines sensitive to $\kappa$-distributions include the He-like \ion{Ne}{IX} lines at 13.55 and 13.71\,\AA, which are however not dealt with in this paper. This is since the \ion{Ne}{IX} lines are not sufficiently numerous to construct a reliable diagnostics of $\kappa$ in the same manner as is done for \ion{Fe}{XVII} and \ion{Fe}{XVIII} in Sect. \ref{Sect:4.2}. We also note that the He-like \ion{Ne}{IX} lines are sensitive to excitation, which singificantly increases their formation temperature compared to the peak of the relative ion abundance. For Maxwellian, the lines are formed at about log($T$\,[K])\,=\,6.6, while the broad peak of the relative ion abundance is at log($T$\,[K])\,=\,6.25. For $\kappa$\,=\,2, the shift is to log($T$\,[K])\,=\,6.5 instead of $\approx$6.1.

\subsection{Theoretical diagnostics}
\label{Sect:4.2}

In the following, we investigate the possible diagnostics of non-Maxwellians using the ratio-ratio diagrams \citep{Dudik14b,Dudik15,Dzifcakova18}. The ratio-ratio diagrams plot the dependence on one line intensity ratio on another line intensity ratio. Typically, one ratio is chosen to be dominantly sensitive to temperature, while the other is sensitive to $\kappa$. Such ratios involve lines separated in wavelength, that is, with different excitation tresholds; and thus sensitive to different parts of the electron energy distribution. Alternatively, a ratio is sensitive to $\kappa$ if the behavior of the collision strengths $\Omega_{ji}$ with energy for the two lines are different. In practice, both ratios in a ratio-ratio diagram are sensitive to both $T$ and $\kappa$ to some degree, since both $T$ and $\kappa$ are independent parameters of the $\kappa$-distribution (see Eq. \ref{Eq:Kappa}).

We use both single-ion ratios as well as combinations of \ion{Fe}{XVII} and \ion{Fe}{XVIII} lines. The list of spectral lines investigated is provided in Appendix \ref{Appendix:Transitions} together with their blends and self-blends. We note that many of the lines are blended (Tables \ref{Table:fe17} and \ref{Table:fe18}), with the importance of some blends, such as those of \ion{Fe}{XIX}, being dependent on the conditions \citep[c.f.,][]{Hutcheon76}.

We also note that the \ion{Fe}{XVII} is not sensitive to electron density, as noted already by \citet{Loulergue75}. This holds to large electron densities, of about 10$^{11}$\,cm$^{-3}$, as also noted by \citet{DelZanna11b}. Above 10$^{11}$\,cm$^{-3}$, some \ion{Fe}{XVII} lines can become density-sensitive, for example, the 17.09\,\AA~line decreases in intensity with respect to the neighboring 17.05\,\AA~one \citep[e.g.,][]{Phillips01}. The \ion{Fe}{XVIII} is similarly not very sensitive to $N_\mathrm{e}$ in the low-density limit relevant for solar corona and flares \citep{DelZanna06}. The validity of this fact can be discerned directly from Figs. \ref{Fig:Diag_single-ion_fe17}--\ref{Fig:Diag_magixs_aia}, where the dash-dotted lines stand for $N_\mathrm{e}$\,=\,10$^{11}$\,cm$^{-3}$, while the full lines denote the spectra calculated for $N_\mathrm{e}$\,=\,10$^{9}$\,cm$^{-3}$. The difference between the two is very small or indeed not recognizable.

\subsubsection{Single-ion diagnostics}
\label{Sect:4.2.1}

We first investigate ratios of lines originating with only a single ion. This has the advantage of not being strongly sensitive to departures from ionization equilibrium \citep{Dudik14b}; although in principle some sensitivity could arise due to the contribution of dielectronic recombination to upper excited levels. However, we focus on lines for which the correction factors (Sect. \ref{Sect:3.3}) are negligible at coronal temperatures. To illustrate the influence of the ionization from and recombination to the excited levels for the Maxwellian distribution, a black dashed line is plotted in Figs. \ref{Fig:Diag_single-ion_fe17}--\ref{Fig:Diag_magixs_aia}. Compared to that, the full black line stands for the Maxwellian spectra without the correction factors. Finally, the colored lines denote individual values of $\kappa$, where we consider $\kappa$\,=\,10, 5, 3, and 2, respectively. 

Unlike in EUV \citep{Dzifcakova13b}, the diagnostic possibilities using only the lines of \ion{Fe}{XVII} are limited. This is primarily due to the fact that the \ion{Fe}{XVII} lines occur only at wavelengths of 10--18\,\AA, and all transitions are to the ground state (Table \ref{Table:fe17}). Most of the line intensity ratios have only a restricted range for $\kappa$\,=\,2 compared to the Maxwellian. An example is shown in the left panel of Fig. \ref{Fig:Diag_single-ion_fe17}. Despite the seeming spread of the ratio-ratio curves for different $\kappa$, the \ion{Fe}{XVII} 13.82\,\AA + 13.89\,\AA~/ 12.26\,\AA~displays only about $\approx$4\% sensitivity to $\kappa$. This is too low for a reliable detection, given the likely photon noise uncertainties in these weak \ion{Fe}{XVII} lines. The conjugate ratio of \ion{Fe}{XVII} 12.26\,\AA\,/\,11.25\,\AA~is however limited to values of about 2.2 (in photon units) for $\kappa$\,=\,2, compared to a much larger range of 2.15--2.60 for the Maxwellian distribution and temperatures of log($T$\,[K])\,=\,6.4--7.0. Therefore, a measurement of this ratio of about 2.2 could be a potential indicator of $\kappa$\,$\approx$\,2.

Ratios of weak \ion{Fe}{XVII} lines with respect to the stronger lines, such as those at 15.01 and 15.26\,\AA, do show sensitivity to $\kappa$. An example is shown in the right panel of Fig. \ref{Fig:Diag_single-ion_fe17}, where the \ion{Fe}{XVII} 15.26\,/\,12.26\,\AA~ratio is combined with the ratio of the \ion{Fe}{XVII} 15.26\,\AA~to the blend of \ion{Fe}{XVII}+\ion{Fe}{XVIII} at 15.45\,\AA. Although the latter ratio involves both \ion{Fe}{XVII} and \ion{Fe}{XVIII}, at temperatures below log($T$\,[K])\,$\leq$\,6.6 (6.8), about 95\% (75\%) of the intensity of the blend is due to a single transition within \ion{Fe}{XVII}. In this ratio-ratio diagram, the sensitivity to $\kappa$ increases toward the lower temperatures, a feature typical of single-ion ratios (c.f., Figs. 8--10 of Paper\,I). Such increase occurs because even for a low temperature, a $\kappa$-distribution with low $\kappa$ still has many high-energy electrons for the excitation to occur. This is advantageous for detection of non-Maxwellians in the plasma of active region cores. 

We note that the 12.26, 15.26, and 15.45\,\AA~lines were well-observed by \citet{Parkinson75} using the KAP (potassium-acid-phtalate) crystal. After transforming the intensities listed in Table 1 therein to photon units, we obtain a value of $\approx$3.0 for the \ion{Fe}{XVII} 15.26\,\AA\,/\,15.45\,\AA~ratio. The conjugate \ion{Fe}{XVII} ratio of 15.26\,\AA\,/\,12.26\,\AA~reaches a value of $\approx$7.5, outside of the range shown in the right panel of Fig. \ref{Fig:Diag_single-ion_fe17}. Such measured ratios would indicate Maxwellian plasma at low temperatures of log($T$\,[K])\,$\lesssim$\,6.3.

Finally, we note that the 15.26\,\AA\,/\,12.26\,\AA~ratio contains contributions level-resolved ionization and recombination (Sect. \ref{Sect:3.3}), which can become important at higher temperatures, see the black dashed line in the right panel of Fig. \ref{Fig:Diag_single-ion_fe17}. Since we are not taking these effects into account, the measured $\kappa$-value for such high temperatures will likely only be an upper limit.

%

The \ion{Fe}{XVIII} produces more lines than \ion{Fe}{XVII} in the wavelength range observed by \textit{MaGIXS}. However, many of these lines are weak compared to the strong \ion{Fe}{XVII} lines at 15.01 and 17.05\,\AA~if the temperatures are log($T$\,[K])\,=\,6.6 typical of active region cores (Fig. \ref{Fig:spec} and Table \ref{Table:fe18}). Indeed, \citet{Parkinson75} did not observe many of the \ion{Fe}{XVIII} lines. In addition to blends, many of the \ion{Fe}{XVIII} lines are self-blended by multiple \ion{Fe}{XVIII} transitions (see Table \ref{Table:fe18}), which can have implications for their widths. Care must thus be exercised in interpreting the intensities of such lines, if observed.

We investigated combinations of the \ion{Fe}{XVIII} line ratios for sensitivity to $\kappa$ and $T$. Similarly to the \ion{Fe}{XVII}, only very weakly sensitive ratios were found, with sensitivities of the order of less than about 5\%. An exception are the ratios involving the 11.42\,\AA~selfblend, which is blended with a \ion{Fe}{XVII} transition at 11.420\,\AA. It is the presence of this blend that creates the sensitivity to both $\kappa$ and $T$, which occurs at low temperatures, log($T$\,[K])\,$\lesssim$\,6.7 (Fig. \ref{Fig:Diag_single-ion_fe18}). In fact, the \ion{Fe}{XVII} blend dominates the line at temperatures below log($T$\,[K])\,$\lesssim$\,6.5. Thus, the ratio-ratio diagrams shown in Fig. \ref{Fig:Diag_single-ion_fe18} involving the 11.42\,\AA~line are no longer single-ion ratios, but involve both \ion{Fe}{XVII} and \ion{Fe}{XVIII}, that is, the ionization equilibrium via the relative ion abundances.

\subsubsection{Diagnostics involving ionization equilibrium}
\label{Sect:4.2.2}

The combination of \ion{Fe}{XVII} lines with one \ion{Fe}{XVIII} line (or vice versa) provides an advantage of utilizing the ionization equilibrium, which is strongly sensitive to temperature, but also sensitive and $\kappa$ (Fig. \ref{Fig:ioneq}, right panel). This ratio of two lines formed in neighboring ionization stages can be combined with another ratio sensitive to $T$ or $\kappa$, leading to a significant increase of sensitivity to departures from the Maxwellian distribution. We note that there are many line ratios available within \ion{Fe}{XVII} and \ion{Fe}{XVIII} to measure the temperature \citep{DelZanna06,DelZanna11b}, although these measurements have been performed so far only under the assumption of a Maxwellian distribution.

Before proceeding to investigate the suitable ratios for simultaneous diagnostics of $T$ and $\kappa$, we caution again that using the ratio-ratio technique involving lines from two neighboring ionization stages can be misleading if the ionization is out of equilibrium, meaning that the relative ion abundances are time-dependent and do not correspond to those shown in Fig. \ref{Fig:ioneq}. Thus, as already noted in Sect. \ref{Sect:3.2}, the diagnostics involving lines from neighboring ionization stages should not be applied if the observed line intensities change on timescales of the order of $\approx$10\,s or less.

Since almost any line intensity ratio involving both \ion{Fe}{XVII} and \ion{Fe}{XVIII} is sensitive to $T$ and $\kappa$, strong lines should be preferentially used to minimize the photon noise uncertainty in the measurements. The strongest lines of \ion{Fe}{XVIII} at 14.20\,\AA, (15.83\,\AA~+ 15.87\,\AA), 16.07\,\AA, and 17.62\,\AA~are recommended. At log($T$\,[K])\,=\,6.6, these \ion{Fe}{XVIII} lines are weak compared to \ion{Fe}{XVII}, having intensities of about 2--10\% relative to the \ion{Fe}{XVII} 15.01\,\AA, depending on the conditions. Nevertheless, these lines should still be measurable, and we note that the \ion{Fe}{XVIII} lines at 14.20\,\AA~and 17.62\,\AA~will have increased intensities for the non-Maxwellians (see Fig. \ref{Fig:spec} and Table \ref{Table:fe18}).

With the exception of 17.62\,\AA, the lines of \ion{Fe}{XVIII} at 14.20\,\AA, 14.37\,\AA, 14.54\,\AA, 15.62\,\AA, (15.83\,\AA~+ 15.87\,\AA), 16.07\,\AA, and 16.17\,\AA~are self-blends. The \ion{Fe}{XVIII} 14.20\,\AA~line is self-blended (hereafter also ``sbl'') with a single 14.21\,\AA~transition (Table \ref{Table:fe18}), which contributes about 38\% to the total intensity of the self-blend. The 14.37\,\AA~and 14.54\,\AA~lines both have many self-blends (Table \ref{Table:fe18}), some of which can contribute tens of per cent to the total intensity. The strongest self-blending transition is 14.55\,\AA, which contributes up to $\approx$30\% to the total intensity of the 14.54\,\AA~line. The 15.62\,\AA~line is self-blended with two transitions at 15.62 and 15.64\,\AA~(Table \ref{Table:fe18}), which however contribute less than about 1\% to the total intensity. The 15.87\,\AA~line is self-blended with another 15.87\,\AA~transition, which contributes about 40\% to the total intensity. The 15.87\,\AA~sbl can be further self-blended at \textit{MaGIXS} resolution with another transition at 15.83\,\AA~(Table \ref{Table:fe18}). In the remainder of this Section, we sum these two lines together unless otherwise indicated. At \textit{MaGIXS} resolution, the 16.07\,\AA~line is located in the red wing of an \ion{O}{VIII}+\ion{Fe}{XVII} blend at 16.00\,\AA~(see Fig. \ref{Fig:spec_log}). This line is self-blended with a 16.09\,\AA~transition, which contributes about 9\% to the total intensity of the self-blend. We note that if hotter temperatures are present, this line can be further blended with a \ion{Fe}{XIX} transition at 16.11\,\AA. Finally, the 16.17\,\AA~line is self-blended with a 16.19\,\AA~transition, which contributes only about 1\% to the total intensity. We note that depending on conditions, perhaps not all of these lines will be reliably measured, especially if shorter integration times are analyzed. Therefore, in the remainder of this work, where \ion{Fe}{XVIII} X-ray lines are concerned, we focus on diagnostics using the strongest lines, such as those at 14.20\,\AA, (15.83\,\AA~+ 15.87\,\AA), 16.07\,\AA, and 17.62\,\AA. Other \ion{Fe}{XVIII} lines could however also be used for consistency checks.

Examples of diagnostic ratio-ratio diagrams using combinations of lines of both \ion{Fe}{XVII} and \ion{Fe}{XVIII} are given in Fig. \ref{Fig:Diag_multi-ion}. There, the \ion{Fe}{XVIII} lines are used in combination with the strong lines of \ion{Fe}{XVII}, such as those at 12.26\,\AA, 15.01\,\AA, 15.26\,\AA, and 16.78\,\AA. We note that the 17.05\,\AA~and 17.09\,\AA~lines, which are formed from a $3s$ upper level, can be used instead of 16.78\,\AA~without a significant change in the shape of the ratio-ratio diagrams. However, the 17.05 and 17.09\,\AA~lines contain larger contribution from correction factors (Sect. \ref{Sect:3.3}); therefore, the use of 16.78\,\AA~line, although a weaker one, is recommended instead. The effect of correction factors is such that the measured value of $\kappa$ will likely be an upper limit once the level-resolved dielectronic recombination is taken into account.

All ratio-ratio diagrams shown in Fig. \ref{Fig:Diag_multi-ion}, involving both \ion{Fe}{XVII} and \ion{Fe}{XVIII} lines observable by \textit{MaGIXS}, have sufficient sensitivity to $\kappa$, of the order of 20\%, for the diagnostics to be feasible. Again, the sensitivity to $\kappa$ is increased at lower $T$, which is advantageous for studying the presence of non-Maxwellians in the active region cores. The best available diagnostics is the \ion{Fe}{XVII} 15.26\,\AA\,/\,\ion{Fe}{XVIII} 17.62\,\AA, strongly sensitive to $T$. This ratio can be combined for example with the \ion{Fe}{XVIII} ratio 17.62\,\AA\,/\,14.20\,\AA~sbl, which is mainly sensitive to $\kappa$ at low temperatures (Fig. \ref{Fig:Diag_multi-ion}). At log($T$\,[K])\,$\lesssim$\,6.6, the latter ratio reaches a value of $\approx$0.31 for $\kappa$\,=\,2 compared to $\approx$0.39 for a Maxwellian.

In the other ratio-ratio diagrams shown in Fig. \ref{Fig:Diag_multi-ion}, the sensitivity to $\kappa$ is chiefly provided by the \ion{Fe}{XVII} ratios such as 15.26\,\AA\,/\,12.26\,\AA, 15.45\,\AA\,(bl \ion{Fe}{XVIII})\,/\,12.26\,\AA, or 16.78\,\AA\,/\,15.01\,\AA. These ratios are combined with the ratios involving both \ion{Fe}{XVII} and \ion{Fe}{XVIII}, such as \ion{Fe}{XVII} 12.26\,\AA\,/\,\ion{Fe}{XVIII} 16.07\,\AA\,sbl, \ion{Fe}{XVII} 15.01\,\AA\,/\,\ion{Fe}{XVIII} 16.07\,\AA\,sbl, or \ion{Fe}{XVII} 16.78\,\AA\,/\,15.45\,\AA~(bl \ion{Fe}{XVIII}). The last case has an added complication such that the \ion{Fe}{XVII} blend with \ion{Fe}{XVIII} at 15.45\,\AA~creates a non-monotonous dependence on $T$ for $\kappa$\,$\gtrsim$\,3 (Fig. \ref{Fig:Diag_multi-ion}, bottom right).

Since these ratio-ratio diagrams involve only \textit{MaGIXS} lines, observed within a single spectral passband of a single instrument, the precision of the diagnostics will be essentially limited by the photon noise and dark current. This a step forward from the former applications of the ratio-ratio diagnostic technique, which involved multiple spectral channels, thereby adding the uncertainty of the absolute calibration of each channel and their degradation over time; see Sect. 4.1 of \citet{Dudik15} and Sect. 4 of \citet{Dzifcakova18}. To take advantage of the photon-noise limited diagnostics, we suggest that the uncertainty in the photon noise should be limited to about 5--10\% for all lines. Since the instrument is likely to observe an active region core, this 5--10\% limit in photon noise uncertainty should be applied to the \ion{Fe}{XVIII} lines, which are weaker under conditions typical of an active region core (c.f., Fig. \ref{Fig:spec}). 

\subsubsection{Diagnostics involving the \ion{Fe}{XVIII} 93.93\,\AA~line}
\label{Sect:4.2.3}

An increase in sensitivity to $\kappa$ can be obtained if the \ion{Fe}{XVIII} 93.93\,\AA~line is used together with the X-ray lines in the ratio-ratio diagrams. Although this well-known line will not observed by \textit{MaGIXS}, it will be simultaneously observed by the \textit{SDO}/AIA instrument in its 94\,\AA~imaging channel. The AIA 94\,\AA~channel is multithermal, but in active region core conditions its signal is dominated by the \ion{Fe}{XVIII} 93.93\,\AA~line \citep[e.g.,][]{ODwyer10,Warren12,DelZanna13b,Petralia14,Brooks16}. The other contributions to this channel, include those at warm coronal temperatures (1--2\,MK); as well as the continuum, which is dependent on both temperature and abundances, are discussed in \citet{DelZanna13b}. The continuum typically contributes 10--20\% to the observed signal. The response of the 94\,\AA~channel of AIA was previously found to be relatively stable over time \citep{Boerner14}. If this stability continues to hold, and if an active region core is observed, the 94\,\AA~channel of AIA can in principle be used as a proxy for the intensity of the \ion{Fe}{XVIII} 93.93\,\AA~line with a sufficient accuracy \citep[$\approx$10\%, depending on the continuum][]{DelZanna13b}.

The line intensity ratio of the 93.93\,\AA~line with respect to X-ray lines of \ion{Fe}{XVIII} is both sensitive to temperature and $\kappa$. Many combinations yield good diagnostic options. Two examples are shown in the top row of Fig. \ref{Fig:Diag_magixs_aia}. The ratios to be preferred are those that involve the strong X-ray lines of \ion{Fe}{XVIII} (see Sect. \ref{Sect:4.2.2}), that is, the lines at 14.20, 15.62, 15.83+15.87, and 17.62\,\AA. The general behavior is such that the sensitivity to $\kappa$ is strongest at low $T$, where it can reach $\approx$30\% of difference between $\kappa$\,=\,2 and Maxwellian at log($T$\,[K])\,=\,6.6.

Much larger sensitivity to $\kappa$ can be obtained if the \ion{Fe}{XVIII} 93.93\,\AA~line is combined with X-ray lines of both \ion{Fe}{XVII} and \ion{Fe}{XVIII}, to take advantage of the behavior of the ionization equilibrium with $\kappa$. Examples of such diagnostics are shown in the bottom row of Fig. \ref{Fig:Diag_magixs_aia}. The sensitivity reached in these cases can be as much as a \textit{factor of $\approx$3} or larger at log($T$\,[K])\,=\,6.6. The ratio-ratio diagram involving the \ion{Fe}{XVII} 15.01\,\AA\,/\,\ion{Fe}{XVIII} 93.93\,\AA~--~\ion{Fe}{XVIII} 93.93\,\AA\,/\ion{Fe}{XVIII} 14.20\,\AA~sbl has both large sensitivity to $T$ and $\kappa$, involves only strong lines, and in good conditions could be used to detect also moderate departures from the Maxwellian. For example, the difference between $\kappa$\,=\,5 and the Maxwellian is still a factor of $\approx$2 at log($T$\,[K])\,=\,6.6. Similar sensitivity is also provided by another of the combinations shown, that of \ion{Fe}{XVII} 16.78\,\AA\,/\,\ion{Fe}{XVII} 15.45\,\AA~(bl \ion{Fe}{XVIII}) -- \ion{Fe}{XVII} 15.45\,\AA~(bl \ion{Fe}{XVIII})\,/\,\ion{Fe}{XVIII} 93.93\,\AA.

\subsection{Diagnostics in the multithermal case}
\label{Sect:4.3}


The diagnostics presented in Sect. \ref{Sect:4} works well for plasma that is isothermal (i.e., at single $T$) and described by a $\kappa$-distribution. However, in many instances, the plasma in active region cores can be multithermal \citep[e.g.,][]{Warren12,Parenti17}. The emerging intensity (Eq. \ref{Eq:line_intensity}) is then given by the expression
\begin{equation}
	I_{ji} = \int A_X G_{X,ji}(T,N_\mathrm{e},\kappa) \mathrm{DEM}_\kappa(T) \mathrm{d}T\,,
	\label{Eq:line_intensity_DEM}
\end{equation}
where the quantity DEM$_\kappa(T)$\,=\,$N_\mathrm{e} N_\mathrm{H} \mathrm{d}l /\mathrm{d}T$ is the differential emission measure \citep{Mason94,Phillips08}. We note that the concept of DEM and the inversion of Eq. (\ref{Eq:line_intensity_DEM}) can be easily extended for the case of $\kappa$-distributions \citep{Mackovjak14,Dudik15,Dzifcakova18} simply by using the contribution functions $G(T,N_\mathrm{e},\kappa)$ for the corresponding value of $\kappa$.

For the case of active region cores, the DEM$_\kappa$ constructed for the $\kappa$-distributions were found to have similar shapes than the Maxwellian ones: \citet{Mackovjak14} found that while the peak of the DEM$_\kappa$ is shifted toward higher $T$ due to the behavior of the ionization equilibrium \citep{Dzifcakova13}, the power-law slope at lower temperatures is nearly the same independently of $\kappa$. The slope of the DEM$_\kappa$ at higher temperatures did change with $\kappa$, but there were only a few lines available to constrain it \citep[][Table 1 and Figs. 2--4 therein]{Mackovjak14}. The \textit{MaGIXS} observations, which contain many high-temperature lines, can be helpful in this regard.

In case the DEM has a pronounced high-$T$ shoulder, the \ion{Fe}{XVIII} lines can be contaminated by blends with \ion{Fe}{XIX} \citep[cf.,][]{Parkinson75}, although in quiescent active region cores the \textit{SMM}/FCS observations at the same wavelengths showed negligible \ion{Fe}{XVIII} and no \ion{Fe}{XIX} \citep{DelZanna14a}. The existence \ion{Fe}{XIX} emission was however observed in active region core conditions in some cases \citep[e.g.,][]{Parenti17}. Thus, the DEM analysis should be performed for the regions observed by \textit{MaGIXS}, for example by using the hotter lines present in the \textit{MaGIXS} spectral channel of 6--24\,\AA. This instrument is well-suited for such analysis, since its spectral band contains lines such as \ion{O}{VII}--\ion{O}{VIII}, \ion{Ne}{IX}--\ion{Ne}{X}, \ion{Mg}{XI}--\ion{Mg}{XII}, and \ion{Fe}{XIX}--\ion{Fe}{XXIV}. Alternatively, the DEM$_\kappa$ can be obtained using the observations made by \textit{SDO}/AIA \citep[see, e.g.,][]{Hannah12,Hannah13,Cheung15} or \textit{Hinode}/EIS \citep[see, e.g.,][]{Warren12,DelZanna13b,Parenti17}.

If the observed plasma is multithermal, the theoretical curves within ratio-ratio diagrams must be folded over the respective DEM$_\kappa$. This has been performed in previous diagnostics of $\kappa$-distributions using this technique \citep{Dudik15,Dzifcakova18}. In essence, the DEM-folding of the ratio-ratio curves is a weighted average, where the weights of individual temperature bins are given by the DEM$_\kappa$. Therefore, the observed intensities for a multithermal plasma will always be \textit{inside} the radius of curvature of the ratio-ratio curves. This is an important effect to take into account, since for majority of cases, the ratio-ratio curves for the $\kappa$-distributions are located toward the inside of the radius of curvature of the corresponding Maxwellian curve. The ratio-ratio diagrams involving the 15.45\,\AA~blend of \ion{Fe}{XVII} and \ion{Fe}{XVIII} (Fig. \ref{Fig:Diag_single-ion_fe17} \textit{right}, Fig. \ref{Fig:Diag_multi-ion} \textit{bottom right}, and Fig. \ref{Fig:Diag_magixs_aia} \textit{bottom left}) are however an exception: For this case, the ratio-ratio curves for the $\kappa$-distributions are \textit{outside} the radius of curvature of the corresponding Maxwellian case. The ratio-ratio diagrams involving this 15.45\,\AA~blend can thus serve as a independent quick check on whether the plasma is multithermal or not, even without the determination of the corresponding DEM$_\kappa$. This is because these ratio-ratio diagrams will yield an apparently different value of $\kappa$ from all the other ratio-ratio diagrams.

%
\section{Summary}
\label{Sect:5}

We investigated the influence of the non-Maxwellian $\kappa$-distributions on the X-ray spectra of \ion{Fe}{XVII} and \ion{Fe}{XVIII} that shall be observed by the \textit{MaGIXS} instrument. \textit{MaGIXS} is an X-ray spectrometer, working at 6--24\,\AA, to be launched in 2020 on a sounding-rocket by NASA. We chose the $\kappa$-distributions, characterized by a power-law, high-energy tail of accelerated electrons, since these distributions have been detected previously in the upper solar atmosphere, from transition region to the corona and flares \citep[e.g.,][]{Dudik15,Dudik17a,Battaglia15,Dzifcakova18,Jeffrey18,Polito18}.

The synthetic \ion{Fe}{XVII}--\ion{Fe}{XVIII} spectra were calculated in ionization equilibrium using the method presented in Paper\,I, which employs the direct integration of the collision strengths over the $\kappa$-distributions. We used atomic data compatible with CHIANTI v8 \citep{DelZanna15a}; however, we did not take into account the corrections to the level populations \citep{Gu03,Landi06} that arise mostly due to contributions from dielectronic recombination. This effect is only a minor one at temperatures corresponding to the active region cores, namely, log($T$\,[K])\,$\approx$\,6.6. More importantly, throughout the temperature range analyzed in this work, the changes in the line intensity ratios due to these corrections are much smaller than the changes due to the $\kappa$-distributions.

At temperatures typical of an active region core, the X-ray spectrum is dominated by \ion{Fe}{XVII} lines. Nevertheless, many of \ion{Fe}{XVIII} lines could be present and visible if some activity is present, even though their intensities will likely not be larger than 10\% of the \ion{Fe}{XVII} 15.01\,\AA~in quiescent conditions. We investigated the \ion{Fe}{XVII} and \ion{Fe}{XVIII} line intensity ratios to be observed by \textit{MaGIXS} for sensitivity to $\kappa$ and $T$. The values of both $T$ and $\kappa$ must be diagnosed simultaneously, since they are both parameters of the electron energy distribution. To do that, we used the ratio-ratio method, where the dependence of one line ratio on another line ratio is evaluated for a range of $\kappa$ and $T$. Multiple combinations of line ratios well-suited for diagnostics of $\kappa$ were found. Typically, the ratio-ratio diagrams involve a ratio of a \ion{Fe}{XVII} line to an \ion{Fe}{XVIII} one, combined with another ratio of two lines formed within a single ion. This type of diagnostics provides sensitivity to $\kappa$ of the order of several tens of per cent. This should be sufficient to diagnose the departures from a Maxwellian using \textit{MaGIXS} spectra. Since the \ion{Fe}{XVII}--\ion{Fe}{XVIII} lines are relatively close in wavelength, the uncertainties in line intensities should be dominated by photon noise, and the uncertainty of the absolute calibration should not play a role.

In addition, the sensitivity to both $\kappa$ and $T$ can be increased if the \textit{MaGIXS} observations are combined with the observations of the well-known \ion{Fe}{XVIII} 93.93\,\AA~line, routinely observed in the 94\,\AA~channel of \textit{SDO}/AIA with a cadence of 12\,s. When this line is used, the line intensity ratios can depart from a Maxwellian by a factor of 2--3 or more, if an extremely non-Maxwellian case of $\kappa$\,=\,2 is present. We note that such cases have been typically detected previously in the solar atmosphere. The large sensitivity afforded by a combination of the \textit{MaGIXS} lines in combination with the AIA 94\,\AA~observations could be more than sufficient to overcome possible difficulties due to cross-calibration of the two instruments.

In summary, the \ion{Fe}{XVII} and \ion{Fe}{XVIII} lines to be observed by the \textit{MaGIXS} instrument are well-suited for detection of the non-Maxwellian distributions in conditions of active region cores, or even at higher temperatures, such as in microflares. Diagnostics of such non-Maxwellian distributions containing accelerated particles would place important constraints on the heating mechanism of the solar corona, as well as provide direct evidence for its impulsive nature.

\begin{acknowledgements}
The authors thank the referee, Dr. Martin Laming, for constructive comments that helped to improve the manuscript. J.D. and E.Dz. acknowledge Grants 17-16447S and 18-09072S of the Grant Agency of the Czech Republic, as well as institutional support RVO:67985815 from the Czech Academy of Sciences. J.D., G.D.Z., and H.E.M. acknowledges support from the Royal Society via the Newton International Alumni Programme. G.D.Z. and H.E.M. also acknowledge support by STFC (UK) via the consolidated grant of the DAMTP atomic astrophysics group (ST/P000665/1) at the University of Cambridge. CHIANTI is a collaborative project involving George Mason University, the University of Michigan (USA), University of Cambridge (UK) and NASA Goddard Space Flight Center (USA).
\end{acknowledgements}


\bibliographystyle{aa}         
\bibliography{Omega_Fe17-18}   

\begin{appendix}

\section{Selected transitions}
\label{Appendix:Transitions}

A list of the spectral lines selected for analysis is provided in Tables \ref{Table:fe17} and \ref{Table:fe18}. Since the calculations outlined in Sect. \ref{Sect:3.2} use hundreds of energy levels, resulting in tens of thousands of transitions, we investigated only the lines that are both located in the X-ray portion of the spectrum, and are strong enough to be observed under conditions typical of an active region core. A line is deemed strong enough if its intensity is at least 0.5\% of the strongest \ion{Fe}{XVII} line, which is the 15.013\,\AA~transition \citep{Parkinson73,Parkinson75,DelZanna11b}.

In addition, the Tables \ref{Table:fe17} and \ref{Table:fe18} also contain information on the self-blending transitions. A transition is considered a self-blend if it is located within $\lesssim$\,0.3\,\AA, and its intensity is at least 1\% of the primary (strongest) transition. Wavelengths of the primary transition within a selfblend are in Tables \ref{Table:fe17} and \ref{Table:fe18} given together with the energy levels involved. Selfblending transitions are indicated together with their level numbers. We note that the amount of self-blending transitions is necessarily a function of the wavelength resolution of the spectrometer. Details on the energy levels can be found in \citet{Liang10b} for \ion{Fe}{XVII} and in \citet{DelZanna06} for \ion{Fe}{XVIII}.

Tables \ref{Table:fe17} and \ref{Table:fe18} also list possible blending transitions. This list have been compiled by using the CHIANTI database to calculate a spectrum corresponding to the default active region DEM, rather than an isothermal spectrum, to obtain a wider range of blends formed at different temperatures. In addition, the Table I of \citet{Parkinson75} was reviewed for any possible unidentified lines present in the vicinity. We note that the Tables \ref{Table:fe17} and \ref{Table:fe18} do not list possible high-temperature blends due to flare lines.

%
%
%
\begin{table*}
\label{Table:fe17}
  \begin{minipage}[t]{\linewidth}
\centering
\renewcommand{\footnoterule}{}  
    \caption[]{Selected \ion{Fe}{XVII} X-ray lines. Strongest transitions are listed together with the wavelengths of their selfblends. Intensities relative to the strongest line are listed for the Maxwellian distribution, $\kappa$\,=\,5 and 2, for log$(T$\,[K])\,=\,6.6 and log($N_\mathrm{e}$\,[cm$^{-3}$])\,=\,9. Possible blending transitions in the vicinity are indicated by ``bl''. ``u'' indicates and unidentified transition. ``th'' denotes that the wavelength is theoretical only. ``w'' denotes a weak line, while ``s'' indicates that the line can be stronger than the corresponding \ion{Fe}{XVII} one.}
    \vspace{-0.8cm}
    \centering
    \renewcommand{\footnoterule}{}  
    \scriptsize
    $$
\begin{tabular}{llclllll}
\hline
\hline
\noalign{\smallskip}
$\lambda~[\AA]$ & levels 	& level description				& $I$(Maxw) 	& $I$($\kappa$=5) & $I$($\kappa$=2) &  selfblends	& notes  		\\
\noalign{\smallskip}
\hline
\noalign{\smallskip}
%
10.504  & 1 -- 181      & $2s^2 2p^6 ~{}^1S_0 - 2s^2 2p^5 7d ~{}^1 P_1 $     	& 0.0095        & 0.0106        & 0.0066        & 10.494 (1--175) 	& bl \ion{Fe}{XVIII} 10.526\,\AA							\\
10.657  & 1 -- 165      & $2s^2 2p^6 ~{}^1S_0 - 2s^2 2p^5 6d ~{}^3D_1 $      	& 0.0089        & 0.0097        & 0.0059        &      			& bl \ion{Fe}{XIX} 10.655\,\AA								\\
10.770  & 1 -- 155      & $2s^2 2p^6 ~{}^1S_0 - 2s^2 2p^5 6d ~{}^1P_1 $      	& 0.0159        & 0.0173        & 0.0106        & 10.767 (1--149)	& bl \ion{Ne}{IX} 10.764\,\AA, \ion{Fe}{XIX} 10.816\,\AA				\\
11.023  & 1 -- 131      & $2s^2 2p^6 ~{}^1S_0 - 2s^1 2p^6 4p ~{}^1P_1 $      	& 0.0060        & 0.0066        & 0.0044        & 11.043 (1--129)	& bl s \ion{Ne}{IX} 11.000\,\AA, \ion{Na}{X} 11.003\,\AA				\\
11.129  & 1 -- 118      & $2s^2 2p^6 ~{}^1S_0 - 2s^2 2p^5 5d ~{}^3D_1 $      	& 0.0328        & 0.0279        & 0.0236        &      			& bl u 11.160\,\AA, bl w \ion{Na}{X} 11.083\,\AA					\\
11.250  & 1 -- 93       & $2s^2 2p^6 ~{}^1S_0 - 2s^2 2p^5 5d ~{}^1P_1 $      	& 0.0329        & 0.0343        & 0.0202        & 11.264 (1--87)	& bl w \ion{Fe}{XVIII} 11.253\,\AA							\\
	&		&							&		&		&		& 11.287 (1--85)	& 											\\
12.124  & 1 -- 71       & $2s^2 2p^6 ~{}^1S_0 - 2s^2 2p^5 4d ~{}^1P_1 $      	& 0.0892        & 0.0879        & 0.0496        &       		& bl \ion{Ne}{X} 12.132\,\AA, 12.137\,\AA~(Ly-$\alpha$)     				\\
12.264  & 1 -- 59       & $2s^2 2p^6 ~{}^1S_0 - 2s^2 2p^5 4d ~{}^3D_1 $      	& 0.0810        & 0.0791        & 0.0443       	&              		& 											\\
12.680  & 1 -- 39       & $2s^2 2p^6 ~{}^1S_0 - 2s^2 2p^5 4s^1 ~{}^1P_1 $      	& 0.0079        & 0.0067        & 0.0032        &              		& bl \ion{Ni}{XIX} 12.656\,\AA 								\\
	&		&							&		&		&		& 			& bl u 12.75\,1\AA~(\ion{Fe}{XVI} d?)					\\
13.159  & 1 -- 37       & $2s^2 2p^6 ~{}^1S_0 - 2s^1 2p^6 3d ~{}^1D_2 $      	& 0.0096        & 0.0089        & 0.0047        &			&											\\
13.825  & 1 -- 33       & $2s^2 2p^6 ~{}^1S_0 - 2s^1 2p^6 3p ~{}^1P_1 $      	& 0.0707        & 0.0685        & 0.0392        &			& bl \ion{Fe}{XIX} 13.844\,\AA~w, bl u 13.868\,\AA					\\
13.890  & 1 -- 31       & $2s^2 2p^6 ~{}^1S_0 - 2s^1 2p^6 3p ~{}^3P_1 $      	& 0.0168        & 0.0140        & 0.0066        &			& bl \ion{Fe}{XIX} 13.872\,\AA~sbl							\\
15.013  & 1 -- 27       & $2s^2 2p^6 ~{}^1S_0 - 2s^2 2p^5 3d ~{}^1P_1 $      	& 1.0000        & 0.8751        & 0.4452        &			& bl w \ion{Fe}{XIX} 14.992\,\AA~sbl, bl u 15.050\,\AA					\\
15.262  & 1 -- 23       & $2s^2 2p^6 ~{}^1S_0 - 2s^2 2p^5 3d ~{}^3D_1 $      	& 0.2929        & 0.2525        & 0.1259        &			& bl w \ion{Fe}{XVI}+\ion{Fe}{XVI}d sbl 15.27\,\AA, bl \ion{Fe}{XIX} 15.208\,\AA, bl u 15.293\AA	\\
15.453  & 1 -- 17       & $2s^2 2p^6 ~{}^1S_0 - 2s^2 2p^5 3d ~{}^3P_1 $      	& 0.0615        & 0.0458        & 0.0184        &			& bl w \ion{Fe}{XVIII} 15.397 sbl,\,15.450,\,15.474,\, 15.508\,\AA			\\
16.004  & 1 -- 14       & $2s^2 2p^6 ~{}^1S_0 - 2s^2 2p^5 3p ~{}^1D_2 $      	& 0.0166        & 0.0129        & 0.0056        &			& bl s \ion{O}{VIII} 16.006\,\AA~sbl (Ly-$\beta$), bl \ion{Fe}{XVIII} 16.005,\,16.026\,\AA		\\
16.238  & 1 -- 10       & $2s^2 2p^6 ~{}^1S_0 - 2s^2 2p^5 3p ~{}^3P_2 $      	& 0.0132        & 0.0101        & 0.0043        &			& bl \ion{Fe}{XIX} 16.272\,\AA								\\
16.336  & 1 -- 7        & $2s^2 2p^6 ~{}^1S_0 - 2s^2 2p^5 3p ~{}^3D_2 $      	& 0.0207        & 0.0158        & 0.0066        &			& bl w \ion{Fe}{XIX} 16.340\,\AA, w \ion{Fe}{XVIII} 16.306\,\AA				\\
16.776  & 1 -- 5        & $2s^2 2p^6 ~{}^1S_0 - 2s^2 2p^5 3s ~{}^3P_1 $      	& 0.7304        & 0.5718        & 0.2472        &			&											\\
17.051  & 1 -- 3        & $2s^2 2p^6 ~{}^1S_0 - 2s^2 2p^5 3s ~{}^1P_1 $      	& 0.9608        & 0.7450        & 0.3184        &			&											\\
17.096  & 1 -- 2        & $2s^2 2p^6 ~{}^1S_0 - 2s^2 2p^5 3s ~{}^3P_2 $      	& 0.8258        & 0.6105        & 0.2418        &			&											\\
\noalign{\smallskip}
\hline
\hline
\end{tabular}
     $$
  \end{minipage}
\end{table*}

%
\begin{table*}[!ht]
\label{Table:fe18}
  \begin{minipage}[t]{\linewidth}
\centering
\renewcommand{\footnoterule}{}  
    \caption[]{Same as for Table \ref{Table:fe17}, but for \ion{Fe}{XVIII}. The line intensities are scaled with respect to the \ion{Fe}{XVII} 15.013\,\AA~line. The 93.93\,\AA~line not observable by \textit{MaGIXS} is added as well.}
    \vspace{-0.8cm}
    \centering
    \renewcommand{\footnoterule}{}  
    \scriptsize
    $$
\begin{tabular}{llclllll}
\hline
\hline
\noalign{\smallskip}
$\lambda~[\AA]$ & levels 	& level description					& $I$(Maxw) 	& $I$($\kappa$=5) & $I$($\kappa$=2) &  selfblends	& notes  								\\
\noalign{\smallskip}
\hline
\noalign{\smallskip}
11.3261 & 1 -- 177      & $ 2s^2 2p^5 ~{}^2 P_{3/2} - 2s^2 2p^4 4d ~{}^2 D_{5/2} $  	& 0.0051        & 0.0114        & 0.0097        & 11.309 (1--178)	& bl \ion{Fe}{XVII} 11.287\,\AA						\\
	&		&								&		&		&		& 11.309 (1--179)	&									\\
	&		&								&		&		& 		& 11.311 (1--174)	& 									\\
	&		&								&		&		& 		& 11.326 (1--173)	& 									\\
	&		&								&		&		& 		& 11.326 (1--175)	& 									\\
11.420  & 1 -- 160      & $ 2s^2 2p^5 ~{}^2 P_{3/2} - 2s^2 2p^4 4d ~{}^2 D_{5/2} $ 	& 0.0048        & 0.0106        & 0.0090        & 11.416 (1--161)	& bl \ion{Fe}{XVII} 11.420\,\AA						\\
	&		&								&		&		&		& 11.420 (1--159)	& 									\\
	&		&								&		&		&		& 11.424 (1--158)	& 									\\
	&		&								&		&		&		& 11.443 (1--155)	& 									\\
	&		&								&		&		&		& 11.446 (1--153)	& 									\\
	&		&								&		&		&		& 11.442 (2--178)	& 									\\
	&		&								&		&		&		& 11.442 (2--179)	& 									\\
	&		&								&		&		&		& 11.446 (2--174)	& 									\\
	&		&								&		&		&		& 11.459 (2--175)	& 									\\
11.525  & 1 -- 137      & $ 2s^2 2p^5 ~{}^2 P_{3/2} - 2s^2 2p^4 4d ~{}^2 F_{5/2} $  	& 0.0053        & 0.0116        & 0.0098        & 11.523 (1--135)	& bl s \ion{Ne}{IX} 11.547\,\AA, bl \ion{Ni}{XIX} 11.539\,\AA	        \\
	&		&								&		&		&		& 11.525 (1--136)	& 									\\
	&		&								&		&		&		& 11.551 (2--161)	& 									\\
	&		&								&		&		&		& 11.562 (2--158)	& 									\\
	&		&								&		&		&		& 11.572 (2--156)	&									\\
13.397	& 1 -- 70       & $ 2s^2 2p^5 ~{}^2 P_{3/2} - 2s 2p^5 3p ~{}^2 D_{5/2} $  	& 0.0034        & 0.0069        & 0.0057        & 13.353 (1--71)	&									\\
	&		&								&		&		&		& 13.374 (1--72)	&									\\
	&		&								&		&		&		& 13.395 (2--80)	&									\\
	&		&								&		&		&		& 13.397 (2--79)	&									\\
	&		&								&		&		&		& 13.424 (1--69)	&									\\
13.962  & 1 -- 60       & $ 2s^2 2p^5 ~{}^2 P_{3/2} - 2s^2 2p^4 3d ~{}^2 D_{5/2} $  	& 0.0039        & 0.0075        & 0.0055        &      			& bl \ion{Fe}{XIX} 13.938\,\AA~sbl, 14.017\,\AA				\\
14.144  & 1 -- 57       & $ 2s^2 2p^5 ~{}^2 P_{3/2} - 2s^2 2p^4 3d ~{}^2 D_{3/2} $  	& 0.0035        & 0.0066        & 0.0047        & 14.124 (2--62)  	&									\\
	&		&								&		&		&		& 14.130 (3--108)	& 									\\
	&		&								&		&		&		& 14.135 (3--107)	& 									\\
	&		&								&		&		&		& 14.136 (1--58)	& 									\\
	&		&								&		&		&		& 14.149 (3--105)	& 									\\
14.204  & 1 -- 56       & $ 2s^2 2p^5 ~{}^2 P_{3/2} - 2s^2 2p^4 3d ~{}^2 D_{5/2} $  	& 0.0673        & 0.1315        & 0.1002        & 14.209 (1--55)	&									\\
14.258  & 1 -- 53       & $ 2s^2 2p^5 ~{}^2 P_{3/2} - 2s^2 2p^4 3d ~{}^2 S_{1/2} $  	& 0.0138        & 0.0262        & 0.0192        & 14.258 (1--52)	&									\\
14.373  & 1 -- 49       & $ 2s^2 2p^5 ~{}^2 P_{3/2} - 2s^2 2p^4 3d ~{}^2 D_{5/2} $  	& 0.0316        & 0.0601        & 0.0442        & 14.344 (2--58)	&									\\
	&		&								&		&		&		& 14.353 (2--57)	& 									\\
	&		&								&		&		&		& 14.395 (1--48)	& 									\\
	&		&								&		&		&		& 14.419 (1--47)	& 									\\
	&		&								&		&		&		& 14.419 (2--55)	& 									\\
14.487  & 1 -- 43       & $ 2s^2 2p^5 ~{}^2 P_{3/2} - 2s^2 2p^4 3d ~{}^4 F_{3/2} $  	& 0.0075        & 0.0134        & 0.0090        & 14.453 (1--46)	& bl \ion{Ni}{XX} 14.470\,\AA						\\
	&		&								&		&		&		& 14.470 (2--53)	& 									\\
	&		&								&		&		&		& 14.477 (1--42)	& 									\\
	&		&								&		&		&		& 14.487 (1--44)	& 									\\
14.537  & 1 -- 41       & $ 2s^2 2p^5 ~{}^2 P_{3/2} - 2s^2 2p^4 3d ~{}^2 F_{5/2} $  	& 0.0252        & 0.0470        & 0.0338        & 14.551 (1--40) 	&									\\
	&		&								&		&		&		& 14.580 (1--37)	& 									\\
	&		&								&		&		&		& 14.580 (3--102)	& 									\\
14.772 	& 3 -- 98       & $ 2s 2p^6 ~{}^2 S_{1/2} - 2s 2p^5 3d ~{}^2 D_{3/2} $    	& 0.0028        & 0.0056        & 0.0042        & 14.760 (3--97)	& bl \ion{Fe}{XIX} 14.738\,\,\AA, \ion{O}{VIII} 14.821\,\,\AA~sbl	\\
	&		&								&		&		&		& 14.772 (2--40)	&									\\
14.890	& 2 -- 33       & $ 2s^2 2p^5 ~{}^2 P_{1/2} - 2s^2 2p^4 3d ~{}^4 D_{3/2} $	& 0.0042        & 0.0069        & 0.0041        & 14.838 (1--29)	& bl \ion{Fe}{XIX} 14.900\,\,\AA~					\\
	&		&								&		&		&		& 14.846 (2--38)	&									\\
	&		&								&		&		&		& 14.868 (2--34)	&									\\
	&		&								&		&		&		& 14.872 (3--89)	&									\\
	&		&								&		&		&		& 14.908 (1--28)	&									\\
15.397 	& 1 -- 15       & $ 2s^2 2p^5 ~{}^2 P_{3/2} - 2s^2 2p^4 3p ~{}^2 D_{5/2} $  	& 0.0042        & 0.0071        & 0.0044        & 15.343 (2--24)	& bl u 15.372\,\AA							\\
	&		&								&		&		&		& 15.369 (1--13)	& 									\\
	&		&								&		&		&		& 15.392 (2--22)	&									\\
	&		&								&		&		&		& 15.398 (1--14)	& 									\\
	&		&								&		&		&		& 15.405 (2--21)	& 									\\
15.450	& 3 -- 77       & $ 2s 2p^6 ~{}^2 S_{1/2} - 2s 2p^5 3s ~{}^2 P_{3/2} $    	& 0.0042        & 0.0073        & 0.0047        & 15.439 (2--19)	& bl s \ion{Fe}{XVII} 15.453\,\AA					\\ 
	&		&								&		&		&		& 15.440 (1--11)	&									\\
	&		&								&		&		&		& 15.450 (3--78)	&									\\
	&		&								&		&		&		& 15.474 (1--12)	&									\\
	&		&								&		&		&		& 15.476 (2--17)	&									\\
	&		&								&		&		&		& 15.508 (2--20)	& 									\\
15.622  & 1 -- 9        & $ 2s^2 2p^5 ~{}^2 P_{3/2} - 2s^2 2p^4 3s ~{}^2 D_{5/2} $	& 0.0263        & 0.0446        & 0.0281        & 15.616 (1--10)	&									\\
	&		&								&		&		&		& 15.644 (2--15)	&		 							\\
15.766  & 1 -- 8        & $ 2s^2 2p^5 ~{}^2 P_{3/2} - 2s^2 2p^4 3s ~{}^2 P_{1/2} $	& 0.0038        & 0.0064        & 0.0040        &      			&									\\
15.828  & 1 -- 7        & $ 2s^2 2p^5 ~{}^2 P_{3/2} - 2s^2 2p^4 3s ~{}^4 P_{3/2} $	& 0.0172        & 0.0285        & 0.0174        &       		&									\\
15.870  & 2 -- 10       & $ 2s^2 2p^5 ~{}^2 P_{1/2} - 2s^2 2p^4 3s ~{}^2 D_{3/2} $	& 0.0178        & 0.0288        & 0.0169        & 15.870 (1--6)		&									\\
16.005  & 1 -- 5        & $ 2s^2 2p^5 ~{}^2 P_{3/2} - 2s^2 2p^4 3s ~{}^2 P_{3/2} $	& 0.0335        & 0.0566        & 0.0355        & 16.026 (2--8)		& bl s \ion{O}{VIII} 16.006\,\AA~sbl (Ly-$\beta$), bl \ion{Fe}{XVII} 16.004\,\AA		\\
	&		&								&		&		&		& 16.026 (3--65)	& 									\\
16.072  & 1 -- 4        & $ 2s^2 2p^5 ~{}^2 P_{3/2} - 2s^2 2p^4 3s ~{}^4 P_{5/2} $	& 0.0479        & 0.0774        & 0.0450        & 16.089 (2--7)		& in wing of \ion{O}{VIII} 16.006\,\AA; bl \ion{Fe}{XIX} 16.110\,\AA	\\
16.166  & 3 -- 64       & $ 2s 2p^6 ~{}^2 S_{1/2} - 2s 2p^5 3s ~{}^2 P_{3/2} $		& 0.0102        & 0.0189        & 0.0129        & 16.183 (3--63)	&									\\
16.306	& 3 -- 61       & $ 2s 2p^6 ~{}^2 S_{1/2} - 2s 2p^5 3s ~{}^4 P_{3/2} $   	& 0.0033        & 0.0060        & 0.0040        &       		& bl s \ion{Fe}{XVII} 16.336\,\,\AA, \ion{Fe}{XIX} 16.340\,\AA       	\\
17.622  & 3 -- 29       & $ 2s 2p^6 ~{}^2 S_{1/2} - 2s^2 2p^4 3p ~{}^2 P_{3/2} $	& 0.0260        & 0.0464        & 0.0306        &       		&									\\
\noalign{\smallskip}
\hline
\noalign{\smallskip}
93.932	& 1 -- 3        & $ 2s^2 2p^5 ~{}^2 P_{3/2} - 2s 2p^6 ~{}^2 S_{1/2} $      	& 0.5790        & 1.0770        & 0.8772        &      			& \textit{SDO}/AIA 94\,\AA~passband					\\
	&		&								&		&		&		&			& bl \ion{Fe}{VIII}, \ion{Fe}{X} s, \ion{Fe}{XII} s, \ion{Fe}{XIV} s, \ion{Fe}{XX}, \ion{Mg}{VIII}	\\
	&		&								&		&		&		&			& see \citet{ODwyer10}, \citet{Warren12}				\\
	&		&								&		&		&		&			& and \citet{DelZanna13b}	\\
\noalign{\smallskip}
\hline
\hline
\end{tabular}
     $$
  \end{minipage}
\end{table*}

%
\section{Photon noise uncertainties in isothermal Maxwellian and $\kappa$\,=\,2 spectra}
\label{Appendix:Spectrum_log}

%
\begin{figure*}
   \centering
   \includegraphics[width=18.6cm,viewport=20 0 850 335,clip]{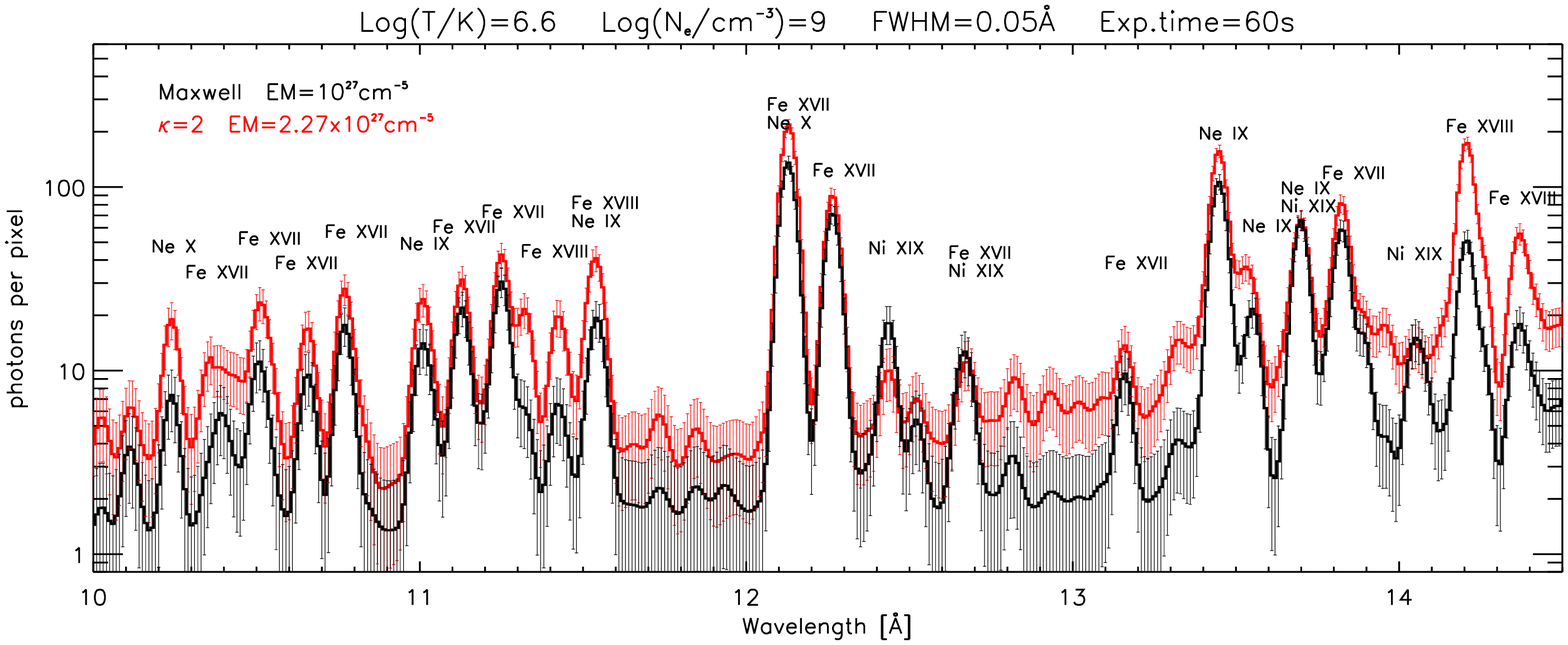}
   \includegraphics[width=18.6cm,viewport=20 0 850 335,clip]{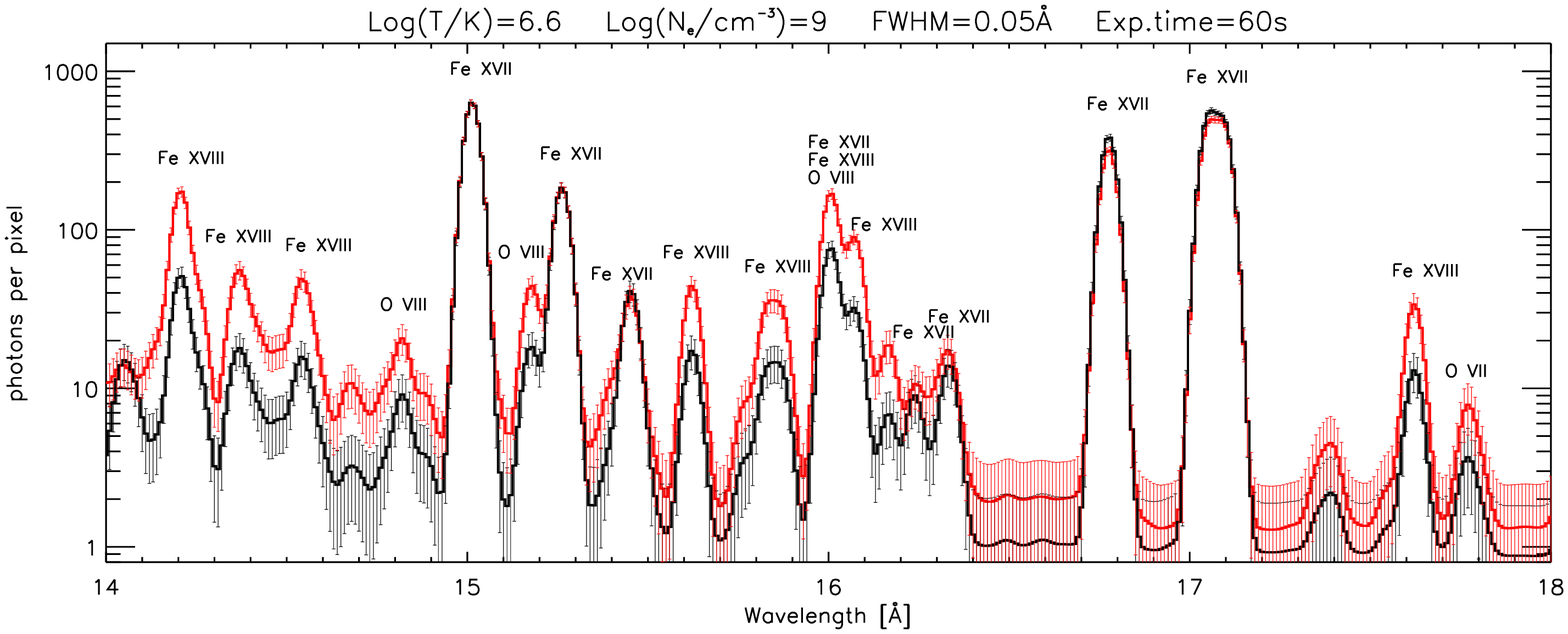}
   \caption{Same as Fig. \ref{Fig:spec}, but with logarithmic intensity axis.}
   \label{Fig:spec_log}%
\end{figure*}
%
%

In Fig. \ref{Fig:spec}, the isothermal spectra at log($T$\,[K])\,=\,6.6 were presented for the Maxwellian and $\kappa$\,=\,2 distributions. As noted in Sect. \ref{Sect:4.1} and also Appendix \ref{Appendix:Transitions}, many of the \ion{Fe}{XVIII} lines can be weak at these temperatures. To further show the photon noise uncertainties in these weaker lines, the spectra of Fig. \ref{Fig:spec} are plotted here using a logarithmic intensity axis (Fig. \ref{Fig:spec_log}).

\end{appendix}

\end{document}